\documentclass[11pt]{article}
\usepackage{latexsym}
\usepackage{amssymb}
\usepackage{epsf}
\usepackage{epsfig}
\usepackage{amsmath}
\usepackage[hypertex]{hyperref}
\usepackage{graphicx}% Include figure files
\usepackage{amscd}
\usepackage{comment}
\usepackage[usenames]{color}

 \makeatletter
    
    \@addtoreset{equation}{section}
  \makeatother

\begin{document}
\pagestyle{empty}

\begin{flushright}
DESY 11-094 \\
KUNS-2340 \\
\today 
\end{flushright}

\vspace{3cm}

\begin{center}

{\bf\LARGE Note on moduli stabilization, \\ supersymmetry breaking and axiverse}
\\

\vspace*{1.5cm}
{\large 
Tetsutaro Higaki\footnote{
E-mail address: tetsutaro.higaki@desy.de}$^{*}$ and Tatsuo Kobayashi\footnote{
E-mail address: kobayash@gauge.scphys.kyoto-u.ac.jp}$^{\dag}$
} \\
\vspace*{0.5cm}

$^{*}${\it DESY Theory Group, Notkestrasse 85, D-22607 Hamburg, Germany}\\
\vspace*{0.5cm}

$^{\dag}${\it Department of Physics, Kyoto University, Kyoto 606-8502, Japan}\\
\vspace*{0.5cm}

\end{center}

\vspace*{1.0cm}

\begin{abstract}
{\normalsize
We study properties of moduli stabilization in 
the four dimensional ${\cal N} = 1$ supergravity theory 
with heavy moduli and would-be saxion-axion multiplets 
including light string-theoretic axions.
We give general formulation for the scenario that 
heavy moduli and saxions are stabilized while axions remain light, 
assuming that moduli 
are stabilized near the supersymmetric solution.
One can find stable vacuum, i.e. non-tachyonic saxions, in the non-supersymmetric Minkowski vacua.
We also discuss the cases, where the moduli are coupled to the
supersymmetry breaking sector and/or moduli have contributions to 
supersymmetry breaking. 
Futhermore we study the models with axions originating from matter-like fields.
Our analysis on moduli stabilization is applicable even if there are not light axion multiplets.}
\end{abstract} 

%%%%%%%%%%%%%%%%%%%%%%%%%%%%%%%%%%%%%%%%%%%%%%%%%%%%%%%%%%%%%%%%%%%%%%%%%%%%
%%%%%%%%%%%%%%%%%%%%%%%%%%%%%%%%%%%%%%%%%%%%%%%%%%%%%%%%%%%%%%%%%%%%%%%%%%%%
\newpage
\baselineskip=18pt
\setcounter{page}{2}
\pagestyle{plain}
\baselineskip=18pt
\pagestyle{plain}

\setcounter{footnote}{0}

%\tableofcontents

\section{Introduction}

Moduli stabilization in superstring theories compactified on the internal space is necessary
to determine 
%each discrete vacua. This means that 
physical parameters such as gauge couplings \cite{Dixon:1990pc}, 
Yukawa couplings \cite{Hamidi:1986vh,Cecotti:2009zf}
and soft supersymmetry (SUSY) breaking parameters \cite{Kaplunovsky:1993rd}
in the visible sector, and to evade the moduli problem \cite{Coughlan:1983ci}
and undesirable new forces \cite{Cicoli:2011yy}.
As a consequence, it also can give several interesting implications to particle physics 
\cite{Choi:2005ge, Falkowski:2005ck, Balasubramanian:2005zx, Acharya:2006ia, Abel:2006qt, BlancoPillado:2004ns},
through the KKLT proposal \cite{Kachru:2003aw} or the racetrack model \cite{Krasnikov:1987jj}.

The complex moduli fields in four dimension typically consist of scalars $\{ \phi \}$ originating from
geometry of compactification space (e.g. its volume) and pseudo-scalars $\{ a \}$ coming from NSNS or RR tensor fields.
Even though all the scalars $\{ \phi \}$ are stabilized,
some of their partners $\{ a \}$ can still remain light due to the
shift symmetries: $a \to a~ +$ const.
Therefore the latter pseudo-scalars are often called string-theoretic axions
\cite{Svrcek:2006yi, Conlon:2006tq, Choi:2006za}
and can include the QCD axion to solve the strong CP problem \cite{Callan:1976je, Peccei:1977hh, Kim:1986ax}\footnote{
If we are to identify one of the axions with the QCD axion, 
the quality of the PQ symmetry needs to be checked for solving the strong CP problem:
$\delta m_{a}^2 \lesssim 10^{-11}(m_{a}^{\rm QCD})^2$. 
Here axion mass $\delta m_{a}^2 $ is a contribution from non-QCD effects,
$m_{a}^{\rm QCD} \approx \Lambda_{\rm QCD}^2/f_a$ is the QCD axion mass just from the instanton,
$f_a$ is the decay constant of the QCD axion and $\Lambda_{\rm QCD} =O(100)$ MeV is the QCD scale.
}.
The number of these axions are originally determined by the topological property of compactified space, 
e.g. the Hodge numbers of Calabi-Yau (CY) three-fold 
\cite{Candelas:1985en}.
%\footnote{
%See for ${\cal N}=1$ supersymmetric effective theories of moduli fields on Calabi-Yau space, 
%for instance \cite{Blumenhagen:2005ga, Grimm:2004uq, Grimm:2004ua}.
%See also for effective field theories including D-branes \cite{Grana:2003ek, Grimm:2008dq, Grimm:2011dx, Denef:2008wq}.
%}
% denoted by $h^{1,1}(X)$ and $h^{2,1}(X)$.
(See also for effective field theories \cite{Blumenhagen:2005ga,Grana:2003ek}.)
Because the numbers can be much larger than of order unity,
one can find many light string-theoretic axions through the moduli stabilization, that is, the string axiverse \cite{Arvanitaki:2009fg}.
The axions can have large axion decay constants beyond the axion window \cite{Janka:1995ir}\footnote{
In the LARGE volume scenario \cite{Balasubramanian:2005zx}, 
one can find $M_{\rm string} \simeq 10^{11} ~{\rm GeV } \ll M_{\rm Pl} = 2.4 \times 10^{18}$ GeV \cite{Conlon:2006tq}.
} and can give influences
on the cosmological observations \cite{Arvanitaki:2009fg}.
%unless the number of axion is not so large and 
%string scale is much smaller than the Planck scale $M_{\rm Pl} = 2.4 \times 10^{18}$ GeV; 
For instance, their misalignment angles and Hubble scale during inflationary epoch are constrained and future observations
of tensor modes and isocurvature perturbations
could suggest the evidence of the (non-)axiverse \cite{Acharya:2010zx}. 
Of course, the relic abundance of the axions should not 
exceed the observed matter density \cite{Komatsu:2010fb}.
This will give interesting constraint not only on the observations but also on the string models in terms of moduli stabilization.
Therefore our purpose is to study general framework 
of moduli stabilization leading to light axions 
based on the ${\cal N}=1$ supergravity (SUGRA).

Besides string-theoretic axions,
one often obtains light field-theoretic axions at low energy, too. 
Thus, in general, the number of axions is estimated as \cite{Higaki:2011bz}
\begin{eqnarray}
\nonumber
 {\mbox{(the number of axions)}} &=& {\mbox{(the number of fields)}} \\
\nonumber
&& +1  
%+{\mbox{(The number of $R$-axion)}}  
- {\mbox{(the number of terms in the $W$)}}.
%{\mbox{(The number of axions)}} = {\mbox{(The number of moduli fields) $+1 -$ (The number of terms in the $W$)}}.
\end{eqnarray}
Here $W$ is the superpotential.
%
%and
%\begin{eqnarray}
%\nonumber
% {\mbox{The number of $R$-axion}} &=& 1  ~~~{\mbox{for $R$-symmetric superpotential}}, \\
%\nonumber
%&=& 0  ~~~{\mbox{when $R$ is explicitly broken.}}
%\end{eqnarray}
%
This is because the Peccei-Quinn (PQ) shift symmetries of fields and
the $R$-symmetry produce candidates of the axions
 whereas independent terms in the superpotential
kill them, assuming the K\"ahler potential $K$ preserves these
symmetries. 
Even if the $R$-symmetry is broken explicitly, this estimate is consistent 
when the constant in the superpotential is involved in the term "the number of terms in the $W$".
Although we have neglected vector multiplets which can become massive, 
they can also reduce the number of axion candidates by absorbing them.
When this counting becomes negative or zero, we do not have any light axions.
If there are very small terms violating PQ
symmetries in $W$ or $K$, they give very light masses to the axions.
%In this paper we will assume $K$ is PQ symmetric, though an instanton effect can break it and give masses to axions.
%We will give a general formula of contribution to axion mass from the
%K\"ahler potential corrections.

In this paper, we study the moduli stabilization scenario leading to 
light axions.
We discuss conditions to give heavy masses to all of real parts of
moduli and leave some of imaginary parts massless.
One of important conditions is SUSY breaking, and 
the typical mass scale is the gravitino mass $m_{3/2}$.
All of the real parts of moduli must have masses, which are larger
than the gravitino mass  
and/or comparable to the gravitino mass.
On the other hand, light axions masses are smaller and 
could be of ${\cal O}(m_{3/2}^{r+1}/M_p^r)$ with $r={\cal O}(1)$ 
or a few tens.

In Section 2, we will study the properties of non-supersymmetric vacua with light string-theoretic axions.
We will also give comments on closed string moduli which are directly coupled to the SUSY breaking sector.
In Section 3, we will study the string-theoretic $R$-axion and the 
saxion-axion multiplet breaking SUSY.
In Section 4, we will discuss corrections to the light axion masses from small breaking terms of PQ symmetries
in the superpotential and the K\"ahler potential.
In Section 5, we will give comments on simple models of field-theoretic axions in terms of effective field theories.
In Section 6, we will conclude this paper. 
Our analysis on moduli stabilization is applicable even if there 
are not  light axion multiplets.
In Appendix, 
several types of moduli stabilization models are briefly reviewed.
We will give a brief comment on the LARGE volume scenario based 
on the recent work of the neutral instanton effect including odd parity moduli under orientifold parity.

\section{Light string-theoretic axions}

In the following sections, we will consider moduli stabilization 
at low energy
with the assumption that irrelevant moduli are heavy by closed string fluxes \cite{Grana:2005jc}.
The remaining moduli of our interest can be stabilized via gaugino condensation \cite{Dine:1985rz}
or (stringy) instanton effects \cite{Witten:1996bn}.
%\footnote{\bf
%With world volume fluxes ${\cal F}$ on the E3-instanton, one can enjoy the {\it neutral} instanton 
%including odd parity moduli under orientifold one \cite{Grimm:2011dj}, 
%e.g. $W = e^{-T-q_{a} G^a-hS}$ in type IIB orientifold with O3/O7,
%where $S$ is the complex dilaton,
%$T$ is the even parity K\"ahler moduli, $G^a$ is the odd parity K\"ahler moduli (or axions) expanded by
%$\omega^a_2 \in H^{1,1}_-(CY)$, $q_a$ and $h$ depend on the ${\cal F}$.
%Even if $T$ and $G$ are charged under anomalous $U(1)$, the superpotential can be neutral 
%in contrast to the paper \cite{Blumenhagen:2007sm} as $G$ always follows $T$.
%This is similar to the Heterotic case \cite{Heteto}.
%Therefore heavy moduli or string-theoretic axion multiplets would correspond not only to $T$ but also to $G$, 
%even though there are necessarily $D$-terms because of $G$.
%This fact can open up new avenues to build realistic models in the string theories.
%}.
%in addition to perturbative effects of $(g_s, \alpha')$.
Thus we study the superpotential below:
\begin{eqnarray}
W &=&  W(\Phi) = W_0+ \sum_k  A_k \exp(-\sum_i a^{(k)}_i \Phi^i)  .
\label{spotential1}
\end{eqnarray}
Here $W_0$ is a constant from the fluxes,
$\{ \Phi^i \}$ are heavy closed string moduli fields which are stabilized 
%not by the fluxes but 
by this superpotential and
we use the unit $M_{\rm Pl} = 2.4 \times 10^{18}$ GeV $\equiv 1$.
We study the possibility that we can have massless axions 
at this stage.
%the perturbative level.
The scalar potential is written by the superpotential $W$ and 
the K\"ahler potential $K$,
\begin{eqnarray}
V &=& V_F \nonumber \\
&=& e^G [G_I G_{\bar{J}}G^{I\bar{J}} -3] \\
&=& e^{K} \left[K^{I\bar{J}}(D_I W)\overline{(D_J W)}-3 |W|^2 \right] 
\nonumber,
\end{eqnarray}
where
\begin{eqnarray}
G= K + \log |W|^2, \qquad D_I W = (\partial_I K) W + \partial_I W .
\end{eqnarray}
Here, 
$K^{I\bar{J}}=G^{I\bar{J}}$ denotes the inverse of 
the K\"ahler metric $K_{I\bar{J}}=\partial_I \bar{\partial}_{\bar{J}} K$.
F-terms and the gravitino mass $m_{3/2}$ are given as 
\begin{eqnarray}
F^I = -e^{G/2}G^I 
= - e^{G/2}G^{I\bar{J}}G_{\bar{J}},~~~
m_{3/2} = e^{G/2}.
\end{eqnarray}
We will focus just on $V_F$ for simplicity.

\subsection{Light string-theoretic axions and saxion masses in the SUSY vacuum}

In this subsection, we briefly review  \cite{Conlon:2006tq}.
We study saxion masses in the SUSY vacuum with light axions.

For instance, let us consider the superpotential with two moduli $(T_1,~T_2)$:
\begin{eqnarray}
W= W_0 + Ae^{-a(T_1 +T_2)} \equiv W_0 + Ae^{-a \Phi}.
\end{eqnarray}
One can find $u \equiv T_1 - T_2$ is absent from the superpotential, that is, we have just one phase of $\Phi$: 
$\partial_u W =0$.
Then the imaginary part Im$(u)$ is a massless axion whereas Re$(u)$
may be stabilized via the K\"ahler potential $K=K(T_i + \overline{T_i})$.
%\footnote{\bf
%In some cases, quantum correction to the K\"ahler potential will be necessary to stabilize all moduli.
%}.

One can generalize this argument to the case with many axions.
Chiral superfields are classified into two classes.
One class of fields $u^\alpha (\equiv \tau^{\alpha} + i b^{\alpha})$ 
do not appear in the superpotential, i.e. 
\begin{eqnarray}
\frac{\partial W}{\partial u^{\alpha}} = 0,
\end{eqnarray}
while the fields $\Phi^i$ in the other class appear.
Then, the imaginary parts of $u^\alpha$, i.e. $b^{\alpha}$ 
are  string-theoretic axions, which have flat directions 
in the scalar potential for the form of K\"ahler potential, 
$K(u + \bar u)$.
We evaluate masses of the real parts of $u^\alpha$, i.e. 
saxions $\tau^\alpha$.
In the SUSY vacuum with stabilized moduli one finds
\begin{eqnarray}
 D_{\hat{i}} W = 0 ~~~{\rm for}~^{\forall}~ \hat{i}=(\Phi^i ,u^\alpha).
\end{eqnarray}
For the fields $u^\alpha$, this leads to 
\begin{eqnarray}
 \frac{\partial K}{\partial u^{\alpha}}  = 0 ~~~{\rm or} ~~~ W=0 .
\end{eqnarray}
In this case, we find 
%\begin{eqnarray}
%\partial_{\tau^{\alpha}} V_F =
%2 e^{K}K^{\hat{i}\bar{\hat{j}}} \left(
%K_{\hat{i}\bar{\alpha}} W \overline{(D_{\hat{j}}W)} + 
%K_{\alpha\bar{\hat{j}}} \overline{W} {D_{\hat{i}}W}
%\right)
%-6 K_{\alpha} e^K |W|^2 +O((DW)^2) ,
%\end{eqnarray}
\begin{eqnarray}
\langle \partial_{\tau^{\alpha}} \partial_{\tau^{\beta}} V_F \rangle _{\rm SUSY} = 4
e^{K} |W|^2 \left[
2 K^{\hat{i}\bar{\hat{j}}} 
K_{\hat{i}(\bar{\alpha}}K_{\beta )\bar{\hat{j}}} 
%+ K_{i\bar{u}^{\beta}}K_{u^{\alpha}\bar{j}} 
-3 K_{\alpha \bar{\beta}}
\right] = -4 e^{K} |W|^2 K_{\alpha \bar{\beta}} \leq 0.
\end{eqnarray}
That is, every massless string-theoretic axion has 
undesirable massless saxion for $W=0$ or tachyonic saxion 
in the SUSY AdS vacuum for $W \neq 0$.
This is because $K_{\alpha \bar{\beta}}$ is the positive definite matrix.
Note that the term $4e^K|W|^2 \cdot (-3 K_{\alpha \bar{\beta}})$ comes from the vacuum energy.
We have used the property of perturbative moduli K\"ahler potential,
\begin{eqnarray}
\partial_{\tau^{\alpha}} K (\Phi+\overline{\Phi};u+\overline{u}) = 2 \partial_{u^{\alpha}} K (\Phi+\overline{\Phi};u+\overline{u}) = 
2 \partial_{\bar{u}^{\alpha}} K (\Phi+\overline{\Phi};u+\overline{u}).
\end{eqnarray}

The tachyonic instability might not be problematic in the AdS vacuum 
because of the Breitenlohner-Freedman bound
\cite{Breitenlohner:1982jf}.
At any rate, one should consider the SUSY breaking Minkowski vacuum 
to realize the realistic vacuum, 
although one may need fine-tuning 
to uplift the SUSY AdS vacuum to the Minkowski one.
Hence, in the following sections, 
we will consider the SUSY breaking effects and then
one can see that the saxions become stable for 
vanishing vacuum energy \footnote{
One can also consider a non-perturbative effect on the K\"ahler potential or
D-term moduli stabilization which means a gauge multiplet eats an axion multiplet to lift saxion direction.
}.

\subsection{Light string-theoretic axions and the saxion mass in the SUSY breaking Minkowski vacuum}

Here, we study saxion stabilizaton in the SUSY breaking 
Minkowski vacuum with light axions.
As a SUSY breaking source, we consider a single 
chiral field $X$.
We assume that moduli $F$-terms $G^{\hat{i}}~(\hat{i} =i,~
\alpha)$ are smaller than $G^X$ and the 
cosmological constant is vanishing, $\langle V_F \rangle = 0$, 
that is,
\begin{eqnarray}\label{eq:GXGX}
G^X G_X \simeq 3,~~~
G^X G_X \gg G^{\hat{i}} G_{\hat{i}} ,
\end{eqnarray}
where $G^A = G^{A\bar{B}}G_{\bar{B}}$. 
%Thus this means moduli are stabilized near the supersymmetric solution.
%In the following discussion, we will assume 
%that ${\cal K} ={\cal K}(\Phi + \Phi^{\dag})$ and $\partial_{\alpha} W =0$.
%

Here, we study the model, where the SUSY breaking sector $X$ and
moduli are decoupled in the K\"ahler potential $K$ 
and the superpotential  $W$.
That is, we consider the following form of 
the K\"ahler potential and the superpotential 
\begin{eqnarray}\label{eq:KW-decouple}
K = \hat{K}(X,\overline{X}) + {\cal K}(\Phi + \overline{\Phi},u+\overline{u}) , ~~~
W = \hat{W}(X) + {\cal W}(\Phi ).
\end{eqnarray}
Hereafter we will set $K_{X \bar{X}} = 1$ at the leading order of
$\overline{X}X$.
Note that $\partial_{\alpha} W =0$ and
$G_{X\bar{\hat{i}}} = K_{X\bar{\hat{i}}} = 0$.
%This is just for simplicity or may be necessary to obtain stable SUSY breaking vacuum.
When there is a large mass splitting between moduli $\Phi$ and $X$, 
$K_{X\bar{i}} \neq 0$ would be possible, 
but $K_{X\bar{i}} \ll 1$ would be necessary for the stable vacuum;
$K_{X\bar{i}} = 0$ would be an appropriate approximation.
A simple example of the SUSY breaking models has $\hat{W}= \mu^2 X$ 
\cite{Dudas:2006gr, Kallosh:2006dv, Abe:2007yb, Abe:2008ka}\footnote{
There are also models including SUSY breaking moduli
\cite{Covi:2008zu}, but we will not consider such models 
since subtle fine-tuning would be necessary.
%On the other hand, if there is not $X$ in the $W$, sequestered
%anti-D3 branes, for instance, can be viable as the SUSY breaking sector.
}. 
At any rate, here we consider generic form of the SUSY breaking 
superpotential $\hat{W}$.

{}From 
the above assumption, one expects moduli $\Phi^i$ and $u^{\alpha}$ are stabilized near the SUSY solution,
\begin{eqnarray}
{\cal K}_i W + {\cal W}_i \sim 0,~~~ {\cal K}_{\alpha} \sim 0,
\end{eqnarray}
such that one obtains heavier moduli masses than 
the gravitino mass $m_{3/2} = e^{G/2}$.
%Otherwise, they will acquire lighter masses through the SUSY breaking effect.
%In the above equation we set $G_{X \bar{X}} =1$.
In the SUSY breaking vacuum with a vanishing cosmological constant, one finds the stationary condition:
\begin{eqnarray}\label{firstd}
\partial_I V_F &=& 
G_I V_F + e^G [
G_I + G^K\nabla_I G_K 
]=0 ,
\end{eqnarray}
which leads to 
\begin{eqnarray}\label{firstd-2}
G_{AI}G_{\bar{B}}G^{A{\bar{B}}} + G_I -G^A G^{\bar{B}} \partial_I G_{A\bar{B}} =0.
\end{eqnarray}
Here $I$ denotes $X,i$, $\alpha$ and
$\nabla$ is a covariant derivative with respect to the K\"ahler metric.
Since $G_{X\bar{i}} = 0$, the above equation becomes
\begin{eqnarray}
% && G_{XX}G_{\bar{X}} + G_{Xi}G_{\bar{j}}G^{i\bar{j}} + G_X - G^X G^{\bar{X}}\partial_X G_{X \bar{X}} \\
&& \sqrt{3}(G_{XX} +1) + G_{X\hat{i}}G_{\bar{j}}G^{\hat{i}\bar{\hat{j}}} 
- G^X G^{\bar{X}}\partial_X G_{X \bar{X}} = 0
~~~{\rm for}~~I=X , \nonumber \\
%&&
%G_{Xi}G_{\bar{X}} + G_{ij}G_{\bar{k}}G^{j\bar{k}} + G_i -
%G^j G^{\bar{k}} \partial_i G_{j\bar{k}} \\
&&  
 \sqrt{3} G_{X\hat{i}} + G_{\hat{i}\hat{j}}G_{\bar{\hat{k}}}G^{\hat{j}\bar{k}} 
+ G_{\hat{i}} - G^{\hat{j}} G^{\bar{\hat{k}}} \partial_{\hat{i}} G_{\hat{j}\bar{\hat{k}}}
=0
~~~{\rm for}~~I=\hat{i} .
\end{eqnarray}
Here, we have used 
\begin{eqnarray}
G_X = G_{\bar{X}}=\sqrt{3},
\end{eqnarray}
because $K_{X \bar X}=1 $ and eq.~(\ref{eq:GXGX}).
Using $G_{X\alpha} =0$, one finds in the vacuum
\begin{eqnarray}
G_{\alpha} = {\cal K}_{\alpha} = \frac{1}{2} G^{\hat{i}} G^{\bar{\hat{j}}} \partial_{\alpha}G_{\hat{i}\bar{\hat{j}}}.
\end{eqnarray}
This means $F$-term of $u^{\alpha}$ is suppressed unless there is mixing between $(X, \Phi^i)$ and $u^{\alpha}$.
%and resembles $D$-term vacuum expectation values (vevs) of anomalous
%$U(1)$ symmetries.
%This would be interesting for $D$-term moduli stabilization.
%and neglected $\partial_X G_{X\bar{X}}$. 
For $X$ and $\Phi^i$, one can typically neglect sub-leading terms
%neglecting sub-leading terms with the assumption that
\begin{eqnarray}
&& G_{X\hat{i}}G_{\bar{\hat{j}}}G^{\hat{i}\bar{\hat{j}}}  \ll 1 ,
\nonumber \\
&& 
G_i - G^{\hat{j}} G^{\bar{\hat{k}}} \partial_i G_{\hat{j}\bar{\hat{k}}}  \ll
\sqrt{3} G_{Xi} + G_{i\hat{j}}G_{\bar{\hat{k}}}G^{\hat{j}\bar{\hat{k}}}  ,
\end{eqnarray}
and one obtains
\begin{eqnarray}
%G_{XX} &\simeq & -1 + {\sqrt{3}} \partial_X G_{X \bar{X}}~~~{\rm or}~~~
\nabla_X G_X \simeq  -1, \qquad 
G^i \simeq - \sqrt{3} (G^{-1})^{ij}G_{X j}  .
%~~~(i,j\neq \alpha ).
\end{eqnarray}
Here $(G^{-1})^{ij}~(i,j\neq \alpha )$ is 
an inverse matrix of $G_{ij} = {\cal K}_{i\bar{j}} + {\cal W}_{ij}/W - {\cal W}_{i}{\cal W}_{j}/W^2$.
%(As $(G^{-1})^{ij}$ is the inverse matrix with $j \neq \alpha$, $i$ does not include $\alpha$, too.)
Thus one can expect the shifts from the SUSY solution of $G^i =0$ and ${\cal K}_{\alpha}=0$ 
are given by
\begin{eqnarray}\label{eq:shift}
\delta \Phi^i \sim {\cal K}_{\bar{k}l}(G^{-1})^{il}(\overline{G}^{-1})^{\bar{k}\bar{j}}G_{\bar{X}\bar{j}},~~~
\delta u^{\alpha} \sim {\cal K}^{\alpha \bar\beta} G^i G^{\bar{j}} \overline{\partial}_{\bar{\beta}} G_{i\bar{j}} 
-{\cal K}^{ \alpha\bar{\beta}}{\cal K}_{\bar{\beta} i}\delta\Phi^i.
\end{eqnarray}
Here we have used typical results
$\sum_{ \bar{k} \supset {\rm heavy~moduli}} {\cal K}_{i{\bar{k}}}{\cal K}^{{\bar{k}j}} \sim \delta_i^j$
and
$\sum_{ \bar{\gamma} \supset {\rm light~moduli}} {\cal K}_{\alpha{\bar{\gamma}}}{\cal K}^{{\bar{\gamma}\beta}} 
\sim \delta_{\alpha}^{\beta}$.
One will see these shifts can be suppressed by the heavy moduli masses squared as $m_{3/2}^2/m^2_{\Phi^i}$.

\subsubsection{Masses for sGoldstino $X$ and heavy moduli $\Phi^i$}

%Suppose that relevant fields other than $u^{\alpha}$s are stabilized. 
We evaluate masses of $X$ and $\Phi^i$.
By differentiating eq.(\ref{firstd}), we obtain in the vacuum
\begin{eqnarray}
\langle V_{I\bar{J}} \rangle & = & e^G [
G_{I\bar{J}} + \nabla_I G_K \bar{\nabla}_{\bar{J}}G^K 
- R_{I\bar{J}K\bar{L}}G^K G^{\bar{L}}
] + (G_{I\bar{J}} - G_I G_{\bar{J}}) V_F , \nonumber \\
\langle V_{IJ} \rangle & = & e^G [
2 \nabla_J G_I + G^K \nabla_J \nabla_I G_K 
] + (\nabla_J G_I -G_I G_J )V_F ,
\end{eqnarray}
where
\begin{eqnarray}
R_{I\bar{J}K\bar{L}} \equiv K_{I\bar{J}K\bar{L}} - K_{IK\bar{A}}K^{\bar{A}B}K_{\bar{J}\bar{L}B} .
\end{eqnarray}
Since we assumed that heavy moduli $\Phi^i$ are stabilized near the SUSY solution, one can neglect $G^I$ term 
to calculate heavy moduli masses $m_{\Phi^i}$ at the leading order of
SUSY breaking effect.

For example, one expects 
\begin{eqnarray}
m_{\Phi^i} \sim a_i \Phi^i m_{3/2} ,
\end{eqnarray}
for the KKLT-like stabilization \cite{Kachru:2003aw} and
\begin{eqnarray}
m_{\Phi^i} \gtrsim (a_i \Phi^i)^2 m_{3/2} ,
\end{eqnarray}
for the racetrack model \cite{Krasnikov:1987jj}, 
which is viable even for $W_0 =0$.
(See also Appendices \ref{app:KKLT} and \ref{app:racetrack} 
for the KKLT-like stabilization and 
the racetrack model, respectively.)
Here $a_i$ denotes the most effective (or smallest) one in $\{ a_i^{(k)} \}$ appearing in the eq.(\ref{spotential1}) 
to the moduli mass $m_{\Phi^i}$.
One could obtain heavier moduli masses than the gravitino mass
by fine-tuning the constant $W_0$ in the racetrack model \cite{Kallosh:2004yh}.

In general, one expects $m_{\Phi^i} \gg m_{3/2}$ and 
mass squared matrix elements of 
the moduli $\Phi$ are written as 
\begin{eqnarray}
V_{i\bar{j}} \simeq e^G [G_{ik}G_{\bar{j}\bar{l}}G^{k\bar{l}}]\equiv
{\cal K}_{i\bar{j}}m_{\Phi^i}^2, 
\qquad V_{ij} \sim 2e^G G_{ij} \equiv 2{\cal K}_{i\bar{j}} m_{3/2}m_{\Phi^i} ,
\end{eqnarray}
that is,
\begin{eqnarray}
V_{i\bar{j}} ~~\gg~~ V_{ij},
\end{eqnarray}
for $m_{\Phi^i} \gg m_{3/2}$.
Note that the mass $\sqrt{V_{i\bar{j}} /{\cal K}_{i\bar{j}}} \simeq 
m_{\Phi^i}$ is the supersymmetric mass of modulus $\Phi^i$.
In the above, we have used the following approximation, 
\begin{eqnarray} 
G_{ij} &=& {\cal K}_{i\bar{j}} + \frac{{\cal W}_{ij}}{W} -\frac{{\cal W}_{i}{\cal W}_j}{W^2}
\simeq {\cal K}_{i\bar{j}} -{\cal K}_i {\cal K}_j + \frac{{\cal W}_{ij}}{W}
\simeq   \frac{{\cal W}_{ij}}{W} \equiv  {\cal K}_{i\bar{j}}
\frac{m_{\Phi^i}}{m_{3/2}} , \nonumber \\
G_{Xi} &=& -\frac{{\cal W}_i \hat{W}_X}{W^2}
\simeq -(\sqrt{3}- \hat{K}_X){\cal K}_i 
\sim  {\cal K}_i \ll G_{ij} , \nonumber \\
G_{ijk} &\sim & \frac{{\cal W}_{ijk}}{W} - \frac{{\cal W}_{ij}{\cal W}_k}{W^2} \sim 
a_k {\cal K}_{i\bar{j}} \frac{m_{\Phi ^i }}{m_{3/2}} + {\cal K}_k {\cal K}_{i\bar{j}} \frac{m_{\Phi ^i }}{m_{3/2}} 
\sim  a_k {\cal K}_{i\bar{j}} \frac{m_{\Phi ^i }}{m_{3/2}}, \nonumber  \\
G_{Xij} &\sim & - \frac{W_{ij}\hat{W}_X}{W^2}
\sim
- {\cal K}_{i\bar{j}} \frac{m_{\Phi^i}}{m_{3/2}} .
\label{contri1}
\end{eqnarray}
We took the diagonal mass matrix $G_{ij}$ for simplifying the discussion here. 
Also one finds
\begin{eqnarray}
G^i \sim  (3 -\sqrt{3}\hat{K}_{X}){\cal K}^{i\bar{j}}{\cal K}_{\bar{j}} \frac{m_{3/2}}{m_{\Phi^i}} 
\sim -(\Phi^i + \overline{\Phi^i}) (3 -\sqrt{3}\hat{K}_{X}) \frac{m_{3/2}}{m_{\Phi^i}}.
\end{eqnarray}
For $G^{\alpha}$ with ${\cal K}_{i \alpha} \neq 0$ for any $i$,
their values are estimated as 
$G^{\alpha} \simeq {\cal K}^{\alpha \bar{i}}\bar{G}_{\bar{i}} \sim G^i$.
Here we have used no-scale like structure 
$\sum_{\bar{j} \supset {\rm heavy ~ moduli}}{\cal K}^{i\bar{j}}{\cal K}_{\bar{j}} \sim -(\Phi^i +\overline{\Phi^i})$ 
up to would-be small perturbative corrections, 
though there is the small $u^{\alpha}$ dependencies ${\cal K}_{\alpha} \sim 0$.
Note that the contribution of eq.(\ref{contri1}) to $V_{ij}$ can be comparable to supersymmetric case, but one still has
$V_{i\bar{j}} \gg V_{ij}$.
Thus, one can obtain the (perturbatively) stable minimum for proper values
of the moduli masses, $m_{\Phi^i}$.
That is, by making $V_{i\bar{j}}$ larger than $V_{ij}$, 
one can realize positive definite mass eigenvalues for all of moduli 
around the SUSY solution $G^i=0$.
Indeed, by using the above result, it is found the shift 
$\delta \Phi^i$ in (\ref{eq:shift}) is suppressed by the factor, 
$m_{3/2}^2/m^2_{\Phi^i}$.

%{\bf By tuning $W$ with a given $K$, one can obtain the (perturbatively) stable minimum. More precisely,
%one can make $V_{IJ}$ smaller than $V_{I\bar{J}}$
%and can get positive definite supersymmetric mass $V_{I\bar{J}}$ via tuning of $G_{IJK}$ and $G_{IJ}$ 
%respectively, except for the sGoldstino obtained by a projection $G_I$ on scalars.
%Here $G_I$ are determined by eq.(\ref{firstd-2}).}

Next, we evaluate the mass of sGoldstino $X$.
The sGoldstino acquires not the mass from $W$ but only SUSY breaking mass from the K\"ahler potential 
because of massless Goldstino in the rigid limit.
There is the necessary condition (not sufficient) for the stable SUSY breaking vacuum, i.e.
non-tachyonic non-holomorphic sGoldstino mass \cite{Covi:2008zu, GomezReino:2006dk}:
\begin{eqnarray}
m^2 &=& V_{I\bar{J}}f^I f^{\bar{J}} =  [3(1+\gamma )\hat{\sigma}
-2\gamma ]m_{3/2}^2 > 0 , 
\end{eqnarray}
where
\begin{eqnarray}
\gamma \equiv \frac{V_F}{3m_{3/2}^2} ,\qquad \hat{\sigma} \equiv
\frac{2}{3} - R_{I\bar{J}K\bar{L}} f^I f^{\bar{J}}f^{K}f^{\bar{L}}, \qquad
f^I \equiv \frac{G^I}{\sqrt{G_K G^K}}.
\end{eqnarray}
For $\gamma=0$ one expects 
\begin{eqnarray}
m^2 &=& 3\hat{\sigma} m^2_{3/2}, \nonumber \\ 
\hat{\sigma} &\simeq & \frac{2}{3} - R_{X\bar{X}X\bar{X}}
= \frac{2}{3} + K_{XX\bar{X}}K^{X\bar{X}}K_{\bar{X}\bar{X}X} - K_{X\bar{X}X\bar{X}} .
\end{eqnarray}
For instance, let us consider the K\"ahler potential with a heavy scale $\Lambda \ll M_{\rm Pl} \equiv 1$
\cite{Kitano:2006wz, Kallosh:2006dv, Abe:2007yb}
\begin{eqnarray}\label{eq:K-corr}
\hat{K} = \overline{X}X - \frac{(\overline{X}X)^2}{4\Lambda^2} + \cdots .
\end{eqnarray}
Then one obtains
\begin{eqnarray}
\hat{\sigma} = \frac{1}{\Lambda^2} + \frac{\overline{X}X }{\Lambda^4} +\frac{2}{3}  > 0.
\end{eqnarray}
Here $\langle X \rangle$ would be of $O(\Lambda^2)$ for the Polonyi model.
For off-diagonal component $V_{XX}$, so long as $G_{XXX} $ and
$\partial_X \Gamma^{X}_{XX} $ are of order unity in the Planck unit, 
one can find $V_{XX} = O(m_{3/2}^2) \ll m^2$.
Thus, there would be the stable minimum.
For string theories, $\Lambda$ would correspond to the mass scale of
heavy field which is coupled to $X$, such as  
anomalous $U(1)$ gauge multiplet mass \cite{ArkaniHamed:1998nu} which is comparable 
to the string scale, when $X$ has the $U(1)$ charge.

\subsubsection{Masses for saxion $\tau^{\alpha}$}

Here, we evaluate masses of saxion $\tau^\alpha$.
One finds {\it positive mass squared}: 
\begin{eqnarray}
%\partial_{\tau^{\alpha}}V_F &=& 2 G_{\alpha}V_F 
%+ 2 e^{G}\left[
%2G_{\alpha} + G_{\hat{i}} G_{\bar{\hat{j}}} \partial_{\alpha}G^{\hat{i}\bar{\ha%t{j}}}
%\right] \\
\langle \partial_{\tau^{\alpha}} \partial_{\tau^{{\beta}}} 
V_F \rangle &=&
4 e^{G} \left[
2 G_{\alpha \bar{\beta}} - \partial_{\hat{i}}G_{\alpha \bar{\beta}}G^{\hat{i}} 
- \bar{\partial}_{\bar{\hat{i}}}G_{\alpha \bar{\beta}}G^{\bar{\hat{i}}} + G_{\hat{i}} G_{\bar{\hat{j}}}
\partial_{\alpha} \bar{\partial}_{\bar{\beta}}G^{\hat{i}\bar{\hat{j}}}
\right] \simeq 8 e^{G}G_{\alpha \bar{\beta}} > 0
.
\label{saxionmass}
\end{eqnarray}
Here we have neglected the last three terms in the bracket, 
since when one obtains $m_{3/2} = e^{G/2} \ll m_{\Phi^i}$
one can find 
\begin{eqnarray}
G^{\hat{i}} \partial_{\hat{i}} G_{\alpha \bar{\beta}} \sim \frac{m_{3/2}}{m_{\Phi}} G_{\alpha \bar{\beta}}.
\end{eqnarray}
Again we have used no-scale like structure ${\cal K}_{\bar{j}}{\cal K}^{i\bar{j}}{\cal K}_{i} = {\rm const.}$
Then the last three terms in eq.(\ref{saxionmass})
are suppressed by $m_{3/2}/m_{\Phi}$ and $(m_{3/2}/m_{\Phi})^2$ respectively, compared to the first term.
%{\bf Note that there are SUSY breaking moduli, i.e. $G^{\Phi}/(\Phi+\overline{\Phi}) \sim 1$, 
%the saxion masses get smaller as we will see later in the Section \ref{axionSB}.}

Instead of $X$, with the sequestered explicit SUSY breaking term $V_{\rm lift} = \epsilon e^{2K/3}$ where 
$\epsilon = 3\langle e^{K/3}|W|^2 \rangle$, one finds the similar results \cite{Choi:2006za},
$\langle \partial_{\tau^{\alpha}} \partial_{\tau^{{\beta}}} 
V_F \rangle \simeq 4 e^{G}G_{\alpha \bar{\beta}}$ and 
$G^i \sim (G^{-1})^{ij}{\cal K}_j \sim (\Phi^i +\overline{\Phi^i})m_{3/2}/m_{\Phi^i}$, i.e. $m_{\alpha} \simeq \sqrt{2}m_{3/2}$.
Here we have neglected the term which is proportional to $K_{\alpha}K_{\beta}$ in 
$\langle \partial_{\tau^{\alpha}} \partial_{\tau^{{\beta}}} 
V_F \rangle$. Note also that mass spectra of heavy moduli for such a
case are similar to ones discussed above.

\subsubsection{Matrix elements}

Here, we summarize the mass matrix.
Including other matrix elements, one can find typically 
\begin{eqnarray}
V_{i\bar{j}} &\sim& e^{G} \left[ G_{ik}G^{k}_{\bar{j}}  + G_{iX}\bar{G}_{\bar{j}\bar{X}} \right]
\sim e^G G_{ik}G^{k}_{\bar{j}} \simeq {\cal K}_{i\bar{j}}m^2_{\Phi^i}
, \nonumber \\
V_{ij} &\sim & e^G \left[ 2 G_{ij} + G_{ijX} \right] \sim 2 e^G G_{ij} \simeq 2{\cal K}_{i\bar{j}} m_{\Phi^i}m_{3/2}
%~<~ V_{i\bar{j}}
,  \nonumber \\
V_{i\bar{X}} &\sim & e^G \left[ G_{ij}G^{j}_{\bar{X}} + G_{iX} \right] 
\sim e^G G_{ij}G^{j}_{\bar{X}} \sim {\cal K}_i m_{\Phi^i}m_{3/2} 
%~< m_{\Phi^i}\sqrt{V_{X\bar{X}}} 
,  \nonumber \\
V_{iX} &\sim & 
%e^G \left[ G_{XXi} + (1 + \Gamma^X_{XX})G_{Xi} + G^j G_{Xij}\right] \sim 
e^G G^j G_{Xij} \sim 
m^2_{3/2} \frac{m_{\Phi^i}}{m_{\Phi^j}} (\Phi^j + \overline{\Phi^{j}}) {\cal K}_{i\bar{j}} \sim
{\cal K}_i m^2_{3/2} ,  \nonumber \\
V_{X\bar{X}} &\sim & - e^G R_{X\bar{X}X\bar{X}}|G^X|^2
\simeq  3\hat{\sigma} m^2_{3/2} , \\
V_{XX} &\sim &e^G \left[ 1 + \nabla_X \nabla_X G_X + G^i ( G_{XXi} + \Gamma^X_{XX}G_{Xi})\right] \sim m^2_{3/2}
%~<~ V_{X\bar{X}}
, \nonumber \\
V_{i\tau^{\alpha}} &\sim &  e^G G_{ki}G^{j}G^{k\bar{m}}{\cal K}_{\alpha j \bar{m}} 
\sim m_{3/2}^2 \frac{m_{\Phi^i}}{m_{\Phi^j}} (\Phi^j + \overline{\Phi^{j}}) {\cal K}_{ij \bar{\alpha}} 
%\sim {\cal K}_{i\bar{\alpha}}m_{3/2}^2 \frac{m_{3/2}}{m_{\Phi}}
\sim {\cal K}_{i \bar{\alpha}} m^2_{3/2} 
, \nonumber
\\
V_{X\tau^{\alpha}} &\sim & 
e^G G_{iX}G^{j}G^{i\bar{m}}{\cal K}_{\alpha j \bar{m}} 
\sim (\Phi^i + \bar{\Phi}^i){\cal K}_{i\bar{\alpha}} m^2_{3/2} \frac{m_{3/2}}{m_{\Phi^i}}
%\sim {\cal K}_{\alpha} m^2_{3/2} \frac{m_{3/2}}{m_{\Phi^j}}
, \nonumber\\
V_{\tau^{\alpha}\tau^{\beta}} &\simeq & 8m^2_{3/2}{\cal K}_{\alpha \bar{\beta}} ,\nonumber
\end{eqnarray}
%
%One can find
%\begin{eqnarray}
%\nonumber
%V_{i\bar{X}} &\sim & {\cal K}_i m_{\Phi^i}m_{3/2},~~~
%V_{iX} \sim 
%e^G \frac{m_{\Phi^i}}{m_{\Phi^j}} (\Phi^j + \overline{\Phi^j}) \partial_j {\cal K}_{\bar{i}}
%\sim {\cal K}_i m^2_{3/2} , \\ 
%V_{i\tau^{\alpha}} &\sim & 
%\sim {\cal K}_{i\bar{\alpha}}m_{3/2}^2 \frac{m_{3/2}}{m_{\Phi}} ,
%m_{3/2}^2 \frac{m_{\Phi^i}}{m_{\Phi^j}} (\Phi^j + \overline{\Phi^j}) \partial_j G_{i \bar{\alpha}} 
%\sim
%G_{i \bar{\alpha}} m^2_{3/2} ,
%~~~
%V_{X\tau^{\alpha}} 
%\sim (\Phi^i + \bar{\Phi}^i){\cal K}_{i\bar{\alpha}} m^2_{3/2} \frac{m_{3/2}}{m_{\Phi^i}}.
%\sim {\cal K}_{\alpha} m^2_{3/2} \frac{m_{3/2}}{m_{\Phi}},
%\end{eqnarray}
%
where we have used
\begin{eqnarray} 
 & & G_{XXi} \sim  
2 \frac{W_X^2 W_i}{W^3} - \frac{W_{XX}W_i}{W^2}
-2 \frac{W_{Xi}W_X}{W^2} 
+ \frac{W_{XXi}}{W} 
\sim 
{\cal K}_i , 
\end{eqnarray}
%\begin{eqnarray}
as well as 
$ \partial_X^n W \lesssim W $.
%\end{eqnarray}
%And we supposed that
%masses of the heavy moduli which are directly coupled to each other in the superpotential
%are the same order of magnitude: 
%\begin{eqnarray}
%m_{\Phi^i} \sim m_{\Phi^j} ~~~{\rm for} ~~W \supset e^{-a_i \Phi^i -a_j \Phi^j}.
%\end{eqnarray}
%
In general, $V_{XX}$ and $V_{i\bar{X}}$ could cause the vacuum instability even if
$m_{\Phi^i} \gg m_{3/2}$ and $\hat{\sigma} > 0$. Based on these matrix elements 
one expects the conditions
\begin{eqnarray}\label{eq:stable}
V_{ij} &<& V_{i\bar{j}}, ~~~~ 
V_{XX} < V_{X\bar{X}} , \nonumber \\
V_{iX},~V_{i\bar{X}} &<& \sqrt{V_{i\bar{j}} V_{X\bar{X}}} , ~~~~
V_{i\tau^{\alpha}} < \sqrt{V_{i\bar{j}} V_{\alpha\bar{\beta}}} , ~~~~
V_{X \tau^{\alpha}} < \sqrt{V_{X\bar{X}} V_{\tau^{\alpha}\tau^{\beta}}} 
\end{eqnarray}
should be satisfied for the (meta)stability.
For this case, so long as $\hat{\sigma} \gg 1$ one would obtain the
stable minimum. 
Then, the mass spectrum is summarized as
\begin{eqnarray}
m^2_{i} \simeq m^2_{\Phi^i} ~ \gg ~ m_{X^{\pm}}^2 \simeq 3\hat{\sigma}m^2_{3/2},~~~
m_{\tau^\alpha}^2 \simeq 4m_{3/2}^2.
\end{eqnarray}
At this stage, the axions $b^\alpha$ are massless.
Note that all of saxions $\tau^\alpha$ corresponding to massless 
axions have almost the same mass $m_{\tau^\alpha} = 2 m_{3/2}$.

Here, after the Goldstino is absorbed into the gravitino, the unnormalized axino masses are given by 
\begin{eqnarray}
(m_{\tilde{a}})_{\alpha \beta} 
%&=& e^{K/2}{\cal D}_i D_j W \\
%&=& e^{G/2}[G_{ij}+G_i G_j -\Gamma^k_{ij}G_k] , \\
%&=& e^{G/2}[\nabla_i G_{j} + G_i G_j ]\\ 
%&\to & 
&=& e^{G/2}\left[\nabla_{\alpha} G_{\beta} + \frac{1}{3}G_{\alpha}
  G_{\beta} \right] \nonumber \\
%\\
%{\cal D}_i D_j W &=& W_{ij} +K_{ij}W +K_i D_j W +K_j D_i W -K_i K_j W -\Gamma^k_{ij}D_k W .
%& \simeq & e^{G/2} \left[G_{\alpha \bar{\beta}} -\Gamma_{\alpha \beta}^i G_i  \right] \\
%& \simeq & e^{G/2} \left[G_{\alpha \bar{\beta}} - \Gamma_{\alpha \beta}^i {\cal K}_i \frac{m_{3/2}}{m_{\Phi^i}} \right]
& \simeq & e^{G/2} G_{\alpha \bar{\beta}} .
\end{eqnarray}
We have neglected $G_{\alpha}$ and $G_i$ because they are of $O(m_{3/2}/m_{\Phi^i})$ corrections.

\subsubsection{$F$-term}

In the above case, one can find
\begin{eqnarray}
F^X \simeq - \sqrt{3}m_{3/2},~~~
\frac{F^{i}}{\Phi^i + \overline{\Phi^i}} \simeq
\sqrt{3}(\sqrt{3} - K_{X}) m_{3/2} \frac{m_{3/2}}{m_{\Phi^i}}  \sim
\frac{F^{\alpha}}{u^{\alpha} + \overline{u^{\alpha}}} 
.
% m_{3/2} \left(\frac{m_{3/2}}{m_{\Phi^i}}\right)^2.
\end{eqnarray}
Here we used the result
%assumed that 
\begin{eqnarray}
G^i &\simeq & G^{i\bar{j}}G_{\bar{j}}, ~~~
G^{\alpha} \simeq G^{\alpha \bar{j}} G_{\bar{j}} ,~~~~
(i,\bar{j} \neq \alpha) , 
\end{eqnarray}
which leads to $G^{\alpha} \sim G^i $.
Even if any $u^{\alpha}$ are stabilized via $D$-terms, $K_{\alpha} \sim 0$,
we gain $F$-term of the $u^{\alpha}$ through the off-diagonal K\"ahler metric \cite{Choi:2011xt, LARGE2, LARGE2-2}.
Note that if $G^{\alpha \bar{j}} = 0$, one finds $F^{\alpha} = 0$ since $G_{\alpha} =0$ for such a case \cite{LARGE3}.
%
%Since $u^{\alpha}$ are light, {\it do we have really such a small $F^{\alpha}$?} 
%Or, is the condition ${\cal K}_{\alpha}=0$ really satisfied?
%
For string-theoretic axion(s) breaking SUSY,
see the Section \ref{axionSB}.

\subsection{Note on mixing between $X$ and moduli and $D$-terms}

For simplicity, 
we have discussed so far the case that the SUSY breaking field $X$ does not couple to moduli $\Phi$ for a simplicity. However,
in string theories, it is natural that moduli are coupled to the SUSY breaking sector via non-perturbative effects,
so that one obtains much smaller scale than the string scale.
Now, let us consider the mixing between $X$ and heavy moduli 
by replacing $\hat W(X)$ in (\ref{eq:KW-decouple}) as follows,
\begin{eqnarray} 
 \hat{W}(X, \Phi) = f(X) \exp[-\sum_i a_i^X \Phi^i_X ] .
\end{eqnarray}
Here $f(X)$ depends only on $X$. 
For instance one can consider the case that $f(X) \sim X$ \cite{Abe:2007yb, Abe:2008ka}
or $f(X) \sim X^{-1}$ \cite{Acharya:2006ia}.
Then, we consider the moduli stabilization with the superpotential,
\begin{eqnarray} 
W =  \hat{W}(X, \Phi) + \sum_k A_k e^{-\sum_i a^{(k)}_i \Phi^i_X} .
\end{eqnarray}
We assume 
\begin{eqnarray} 
a^X_i \sim a_i ,
\end{eqnarray}
in the above superpotential, where  
 $a_i$ is the most effective one to the moduli mass in $a^{(k)}_i$ for $\Phi^i_X$.
Then one can find
\begin{eqnarray}
W_{Xi} \sim - a_i W_{X} \simeq -  a_i (\sqrt{3}-K_X) W .
\end{eqnarray}
Also, one obtains for $\Phi^i_X$
\begin{eqnarray} 
G_{Xi} &=& \frac{W_{Xi}}{W} -\frac{W_i W_X}{W^2}
\simeq 
-(\sqrt{3}-K_X)(a^X_i + {\cal K}_i ) 
%\sim  (a_i + {\cal K}_i ) 
\lesssim G_{ij}  , \nonumber \\
G_{Xij} &\sim & \frac{W_{Xij}}{W} - \frac{W_{ij}W_X}{W}
\sim
a^X_i a^X_j - {\cal K}_{i\bar{j}} \frac{m_{\Phi^i}}{m_{3/2}} 
\sim  a^X_i a^X_j \gtrsim G_{ij} , \\
G_{XXi} &\sim & 
2 \frac{W_X^2 W_i}{W^3} - \frac{W_{XX}W_i}{W^2}
-2 \frac{W_{Xi}W_X}{W^2} 
+ \frac{W_{XXi}}{W} 
\sim 
{\cal K}_i
+ a^X_i 
%\sim 
%{\cal K}_i
%+ a_i 
,\nonumber
\end{eqnarray}
and  also we estimate 
\begin{eqnarray} 
G^i & \sim &(G^{-1})^{ij}G_{jX} \sim {\cal K}^{i\bar{j}}a^X_j \frac{m_{3/2}}{m_{\Phi^i}}
\sim a (\Phi+\bar{\Phi})^2 \frac{m_{3/2}}{m_{\Phi}}, \nonumber\\
G^{\alpha} &\sim & {\cal K}^{\alpha \bar{i}}\bar{G}_{\bar{i}}
\sim G^i .
\end{eqnarray}
For metastability,
one expects the conditions (\ref{eq:stable})
%\begin{eqnarray}
%V_{ij} &<& V_{i\bar{j}}, ~~~~ 
%V_{XX} < V_{X\bar{X}} , \nonumber\\
%V_{iX},~V_{i\bar{X}} &<& \sqrt{V_{i\bar{j}} V_{X\bar{X}}} , ~~~~
%V_{i\tau^{\alpha}} < \sqrt{V_{i\bar{j}} V_{\tau^{\alpha}\tau^{\beta}}} , ~~~~
%V_{X \tau^{\alpha}} < \sqrt{V_{X\bar{X}} V_{\tau^{\alpha}\tau^{\beta}}} .
%\end{eqnarray}
should be satisfied.

%In the appendix \ref{mme2}, we showed matrix elements.
Here, with the assumption that $G_{XXX} =O(1)$, one finds for $\Phi^i_X$
\begin{eqnarray}
V_{i\bar{j}} &\sim& e^{G} \left[ G_{ik}G^{k}_{\bar{j}}  + G_{iX}\bar{G}_{\bar{j}\bar{X}} \right]
%\sim e^G G_{ik}G^{k}_{\bar{j}} 
\sim {\cal K}_{i\bar{j}}m^2_{\Phi^i} + a^X_i a^X_j m_{3/2}^2
 , \nonumber \\
V_{ij} &\sim & e^G \left[ 2G_{ij} + G_{ijX} + G^k G_{ijk}\right] 
%\sim 2e^G G_{ij} 
\sim {\cal K}_{i\bar{j}} m_{\Phi^i}m_{3/2} + a^X_i a^X_j m^2_{3/2} + a^X_i a_j m_{3/2}^2\frac{m_{\Phi^i}}{m_{\Phi^j}}
%~<~ V_{i\bar{j}}
, \nonumber \\
V_{i\bar{X}} &\sim & e^G 
\left[
G_{ij}G^{j}_{\bar{X}} 
+ G_{iX} \right] 
%\sim e^G G_{ij}G^{j}_{\bar{X}} 
\sim a^X_i m_{\Phi^i}m_{3/2}  + a^X_i m_{3/2}^2
%~< m_{\Phi^i}\sqrt{V_{X\bar{X}}} 
, \nonumber \\
V_{iX} &\sim & e^G \left[ G_{XXi} + (1 + \Gamma^X_{XX})G_{Xi} + G^j G_{Xij}\right] 
%\sim e^G G^j G_{Xij} 
%&\sim & e^G G^j G_{Xij} 
\sim  a^X_i {\cal K}^{j\bar{k}}a^X_j a^X_k  m^2_{3/2}\frac{m_{3/2}}{m_{\Phi^j}} + a_i^X m_{3/2}^2
, \nonumber \\
V_{X\bar{X}} &\sim & G_{iX}G^{i}_{\bar{X}} - e^G R_{X\bar{X}X\bar{X}}|G^X|^2
\simeq 3\hat{\sigma}_{R} m^2_{3/2} + {\cal K}^{i\bar{j}}a^X_i a^X_j m_{3/2}^2
, \nonumber \\
V_{XX} &\sim & e^G \left[ 1 + \nabla_X \nabla_X G_X + G^i ( G_{XXi} + \Gamma^X_{XX}G_{Xi})\right] 
\sim m^2_{3/2} \left( 1 + {\cal K}^{i\bar{j}}a^X_i a^X_j \frac{m_{3/2}}{m_{\Phi^i}} \right), \nonumber \\
%~<~ V_{X\bar{X}}.
V_{i\tau^{\alpha}} &\sim &
e^G \left[ 
G_{i\bar{\alpha}} +
G_{ki}G^{j}G^{k\bar{m}}{\cal K}_{\alpha j \bar{m}} 
\right] 
%\sim {\cal K}_{i\bar{\alpha}} a (\Phi +\bar{\Phi}) m^2_{3/2} \frac{m_{3/2}}{m_{\Phi}} 
\sim {\cal K}^{j\bar{k}}{\cal K}_{ij\bar{\alpha}} a^X_k m^2_{3/2} \frac{m_{\Phi^i}}{m_{\Phi^j}} 
+ {\cal K}_{i\bar{\alpha}} m_{3/2}^2
\nonumber \\
&\sim & {\cal K}^{j\bar{k}}{\cal K}_{ij\bar{\alpha}} a^X_k m^2_{3/2} + {\cal K}_{i\bar{\alpha}} m_{3/2}^2
, \nonumber  \\
V_{X\tau^{\alpha}} &\sim &
e^G G_{iX}G^{j}G^{i\bar{m}}{\cal K}_{\alpha j \bar{m}} 
\sim {\cal K}^{i\bar{l}}a^X_i {\cal K}^{j\bar{m}}a^X_m {\cal K}_{j\bar{l}\alpha} m^2_{3/2} \frac{m_{3/2}}{m_{\Phi^j}}
%\sim {\cal K}^{i\bar{j}}a_i a_j {\cal K}_{\alpha} m^2_{3/2} \frac{m_{3/2}}{m_{\Phi^j}}
, \nonumber \\
V_{\tau^{\alpha}\tau^{\beta}} &\sim &
{\cal K}_{\alpha \bar{\beta}}  m^2_{3/2}(8 + a^X_i \Phi^i \frac{m_{3/2}}{m_{\Phi^i}} ) 
\sim {\cal K}_{\alpha \bar{\beta}}  m^2_{3/2} ,
\end{eqnarray}
where $\hat{\sigma}_R$ denotes only $R_{X\bar{X}X\bar{X}}$ contribution in $\hat{\sigma}$.

However, if the linear combination of $a_i^X\Phi_X^i$ were stabilized
via 
a KKLT-like model, 
i.e. $DW|_{\rm KKLT} \sim 0$ and $m_i \sim (a_i\Phi^i_X)m_{3/2}$,
one would obtain 
\begin{eqnarray}
G^X G_X \sim G^i G_i ,
\end{eqnarray}
in addition to $V_{iX} \sim V_{i\bar{X}} \sim  \sqrt{V_{i\bar{j}} V_{X\bar{X}}}$ for $\hat{\sigma}_R \lesssim a_i^X \Phi^i_X$,
$V_{i\bar{j}} \sim V_{ij}$ and $V_{i\tau^{\alpha}} \sim \sqrt{V_{i\bar{j}} V_{\tau^{\alpha}\tau^{\beta}}}$.
This means the assumption that $G^X$ is the main source of the SUSY breaking is violated;
KKLT stabilization of $a_i^X\Phi_X^i$ and realization of the Minkowski
vacuum can not be realized successfully  
and the vacuum would be destabilized to the SUSY AdS one \cite{Abe:2007yb, Dudas:2008qf, Krippendorf:2009zza}.
Even if the assumption that $a_i \sim a_i^X$ is violated, 
the uplifting to the Minkowski vacuum with KKLT stabilization of $a_i^X\Phi_X^i$ would fail 
since there would be the runaway direction, e.g., for small $X$.
%
%(Small $X$ vev could be dangerous for some cases.)
%One can conclude when the linear combination of moduli $a_i^X \Phi_X^i$ 
%is stabilized via KKLT-like model, one always needs the uplifting sector 
%which depends only on the heavier moduli, for instance \cite{Dudas:2007nz, Dudas:2008qf}
%\begin{eqnarray}
%W &=& W_0 + e^{-a\Phi}X + M X X' ~~~ \to ~~~
%W = W_0 + e^{-a\Phi'} + \mu^2 X',  \\
%\partial_{\Phi} M &=& 0, ~~~
%\Phi' = \Phi -\frac{1}{a}\log(X) , ~~~\mu^2 = M \langle X \rangle .
%\end{eqnarray}
%Here we have assumed $X$ develops vev and becomes massive, e.g., via anomalous $U(1)$ symmetry.
Thus, the linear combination of moduli $a_i^X \Phi_X^i$, 
which are coupled to the SUSY breaking sector $X$, should be stabilized 
via racetrack model \cite{Acharya:2006ia, Abe:2007yb, Abe:2008ka}\footnote{
For racetrack stabilization of $\Phi_X^i$, the condition that
$V_{i\bar{X}} <  \sqrt{V_{i\bar{j}} V_{X\bar{X}}}$
would be subtle for $\hat{\sigma}_R \lesssim a_i^X \Phi^i_X$. 
However, one can find the stable vacuum in the concrete models. 
}, fluxes, or D-terms \cite{Heckman:2008qt, Choi:2011xt},
so that they gain much heavier masses than the KKLT-type mass, 
$m_i \gtrsim (a_i\Phi^i_X)^2m_{3/2} \gg (a_i\Phi^i_X) m_{3/2} \gg m_{3/2}$.
(See also \cite{Higaki:2011bz} for models in which there is the coupling between the SUSY breaking sector and the saxion-axion multiplet. In the model, one finds also the saxion mass much larger than the gravitino mass via the K\"ahler stabilization.)

For D-term stabilization $\partial_{\Phi^i}{\cal K}=0$, the moduli 
charged under anomalous $U(1)$ symmetries can become massive
by U(1) symmetry breaking and the massive vector multiplet's eating them, 
even though $\partial_{\Phi^i} W =0$ if matter vevs become consequently irrelevant to the vector mass $M_V$:
%With the assumption that matter vevs are negligible 
%(due to the signature of the charges or very small FI term),
%D-term scalar potential can be written by 
%\begin{eqnarray}
%V_D &=& \frac{{\rm Re}(f)}{2}D^2 = \frac{1}{2{\rm Re}(f)} (\eta^i{\cal K}_i)^2
%= \frac{1}{2{\rm Re}(f)} (\eta^{\Phi_X}{\cal K}_{\Phi_X})^2 ,
%\end{eqnarray}
%and  the heavy saxion and gauge boson mass $M_V$ are obtained as 
%\begin{eqnarray}
%&& \partial_{\Phi_X}^2 V_{D} 
%= \frac{1}{{\rm Re}(f)}(\eta^i {\cal K}_{i \Phi_X})^2 = 
%\frac{1}{{\rm Re}(f)}(\eta^{\Phi_X} {\cal K}_{\Phi_X \overline{\Phi_X}})^2, 
%\end{eqnarray}
%which is comparable to the string scale
\begin{eqnarray}
m^2_{\Phi_X} \equiv M_V^2 \simeq g^2 \eta^{\Phi_X} \eta^{\Phi_X}{\cal K}_{\Phi_X \overline{\Phi_X}} .
\end{eqnarray}
%This is a correction to $V_{i\bar{j}}$ and $V_{ij}$.
%Here we assumed $\partial_{\Phi_X} {\cal K} \geq 0$, $a>0$ and $q_X = \eta^{\Phi_X}/a >0$, 
%and then neglected $X$ vev.
%
Here $\eta^{\Phi_X}$ is the variation of $\Phi_X$ under the anomalous $U(1)$ and $M_V$ from $\Phi_X$ 
can be comparable to the string scale.
Thus for such a case, one can find SUSY breaking Minkowski vacuum, i.e.
%For instance, the superpotential  $W = W_0 + e^{-a\Phi}X$ 
%can be replaced by 
%\cite{Heckman:2008qt, Choi:2011xt},
%\begin{eqnarray}
%W = W_0 + \mu^2 X, ~~~ \mu^2 =  e^{-a\langle \Phi \rangle} .
%\end{eqnarray}
%Here we have assumed that 
%$\Phi$ develops (vanishing) vev and becomes massive via the anomalous
%$U(1)$ symmetry breaking.
%
%Our computation would be plausible even with $D$-terms
%when one integrates out heavy matter and heavy moduli absorbed into gauge boson multiplets
%in front of the non-perturbative superpotential via $F$-term or
%$D$-term condition \cite{Choi:2006bh}.
via F-term \cite{Choi:2006bh} or D-term conditions \cite{Heckman:2008qt, Choi:2011xt},
the superpotential  
$W \sim A_0 (\Psi)e^{-a_i\Phi^i} + e^{-a^X_i\Phi_X ^i}X $ can be replaced by 
\begin{eqnarray}
W \sim  A_0(\langle \Psi \rangle )e^{-a_i\Phi^i} + e^{-a_i^X \langle \Phi_X^i \rangle }X 
\equiv A e^{-a_i\Phi^i} + \mu^2 X 
\end{eqnarray}
in the low energy limit. Here $\{\Psi \}$ are open string modes.
%with $D \sim 0$.
%At any rate, when the SUSY breaking source and moduli are mixing 
%in the superpotential, we need a mechanism to make such moduli heavy, 
%e.g. the D-term and the racetrack model.
In the paper \cite{Choi:2011xt}, when one obtains the tiny Fayet-Iliopoulos term
\begin{eqnarray}
\frac{M_V^4}{3} \simeq {\xi_{\rm FI}}  
\end{eqnarray}
so that $\Phi_X$ is absorbed into vector multiplet,
one can find Minkowski vacuum due to the Polonyi model in the low energy limit.
Here $\xi_{\rm FI} = \eta^{\Phi_X}\partial_{\Phi_X} {\cal K}$ 
is the Fayet-Iliopoulos term from moduli $\Phi_X$.
For such a case, $F^{\Phi_X} \sim \eta^{\Phi_X}m_{3/2} \sim 10^{-2}m_{3/2}$ is obtained with D-term stabilization.
(Note that one may find $M_V^2 \ll {\xi_{\rm FI}} $ if 
$\partial_{\Phi_X}^2 {\cal K} \sim \partial_{\Phi_X} {\cal K}$ and $\eta^{\Phi_X} \ll 1$.)

\section{Approximate $R$-symmetry, $R$-axion and SUSY breaking moduli}

In this section, we study the model, which has an approximate 
$R$-symmetry and $R$-axion.
We also study the model, where SUSY is also breaking by moduli fields.
Indeed, we show that both models are investigated 
in the same way.

\subsection{$R$-axion and SUSY breaking moduli}

In general, a global $U(1)$ $R$-symmetry is broken explicitly because string theory describes the quantum gravity.
Indeed, string models with the exact and global $U(1)$ $R$ symmetry
have not been found.
For instance the constant $W_0$ in the superpotential is easily obtained via flux compactifications, 
but the value depends on the choice of the flux vacua \cite{Bousso:2000xa}.
Therefore at a certain scale there may be an approximate $R$-symmetry accidentally in the SUSY breaking sector and
the moduli stabilization sector
when one obtains $W_0 = 0$ in the superpotential \cite{Kachru:2002he}.

For example, the following superpotential, 
\begin{eqnarray}
W= Ae^{-a \Phi} ,
\end{eqnarray}
has the R-symmetry, where the field $\Phi$ 
transforms as $\Phi \to \Phi - i\frac{2}{a}\alpha$ 
under the R-transformation with a transformation parameter $\alpha$.
Similarly, the racetrack model has the R-symmetry 
\cite{Abe:2005pi, Abe:2007yb, Abe:2007ax, Acharya:2006ia} 
if one has more than two fields in the superpotential without $W_0$.
Thus, when ${\rm Re}(\Phi)$ is stabilized by the K\"ahler potential
for example, we obtain the so-called light $R$-axion.

Here, we consider the R-symmetric superpotential.
%From now on, we will consider the $R$-axion and
%suppose that one can assign the $R$-symmetry.
Then one can rewrite the superpotential including SUSY breaking sector $X$,
\begin{eqnarray}
W &=& e^{-{\cal R}}{\cal W}(X,\Phi), 
\end{eqnarray}
where $\partial_{{\cal R}} {\cal W} = 0$.
Since ${\cal R}$ can include not only $X$ but also moduli in the linear combination, we call it string-theoretic $R$-axion. 
Only ${\cal R}$ transforms as ${\cal R} \to {\cal R} + i2\alpha$ 
under the $R$-symmetry, while the others do not transform.
Note that by the K\"ahler transformation with holomorphic function ${\cal G}$, 
\begin{eqnarray}
K &\to & K + {\cal G} + \bar{{\cal G}},~~
W \to \exp[-{\cal G}]W ,~~~  G \to G ,
\end{eqnarray}
physics is invariant since the action is written by only the total K\"ahler potential $G= K +\log |W|^2$.
Thus one can consider the following K\"ahler potential $K$ and 
the superpotential $W$,
\begin{eqnarray}
K = K^{(0)} - ( {\cal R} + \bar{{\cal R}} ),~~~ W= {\cal W} .
\end{eqnarray}
Here $K^{(0)}$ is the original K\"ahler potential obtained from the dimensional reduction.
Then one finds
\begin{eqnarray}
G_{{\cal R}} = K_{{\cal R}} = K^{(0)}_{{\cal R}} - 1 ,~~~G_I = K_I + \frac{{\cal W}_I}{{\cal W}}
~~~{\rm for}~I \neq {\cal R}.
\end{eqnarray}
Hence unless $G_{{\cal R}}=0$, the $R$-axion is a source of the SUSY breaking.
By Nelson and Seiberg argument \cite{Nelson:1993nf, Abe:2007ax}, 
the existence of the $R$-axion means the SUSY breaking,
provided the model is generic and calculable.
Hence we will also consider the SUSY breaking moduli with the vanishing cosmological constant: 
$G_{\cal R} \neq 0$ and $\langle V_F \rangle =0$.

Because the differences between string-theoretic $R$-axion and string-theoretic axions $u$
are just that the K\"ahler potential and their first derivatives as we saw,
the following results are applicable not only to the string-theoretic $R$-axion, 
but also to usual string-theoretic axions $u$, 
which have non-trivial contributions to SUSY breaking.
%{\bf  
%Hence, if one considers the following K\"ahler potential with $G_{{\cal R}}=0$:
%$K = \hat{K}(X,\overline{X}) + {\cal K}(\Phi+\overline{\Phi};{\cal R}+\overline{\cal R})$,
%one can find the same results as those of the string-theoretic axions $u$.
%Therefore we will focus on the SUSY breaking case $G_{{\cal R}} \neq 0$.
%}

\subsection{SUSY breaking string-theoretic ($R$-)axions}
\label{axionSB}

Let us consider the K\"ahler potential 
\begin{eqnarray}
K = \hat{K}(X, \overline{X}) + \tilde{K}({\cal R}+\overline{{\cal R}}) + {\cal K}(\Phi+\overline{\Phi}) ,
\end{eqnarray}
with $G_{{\cal R}}\neq 0$.
%which never dominates over the SUSY breaking $G^{{\cal R}}G_{{\cal R}} \lesssim G^X G_X$.
For simplicity we will study the case that the K\"ahler potential is separable
and focus only on the SUSY breaking string-theoretic ($R$-)axion 
neglecting dynamics of heavy moduli $\Phi$.
Note that the discussion in this section is applicable 
to an usual string-theoretic axion $u$, which have non-trivial 
contributions to SUSY breaking.

One obtains the stationary condition with the vanishing cosmological constant:
\begin{eqnarray}
\nabla_X G_X & \simeq & -1 , \nonumber \\ 
G^{{\cal R}} &=& \frac{G_{{\cal R}}}{\tilde{K}_{{\cal R}\bar{\cal R}}}
\simeq
2 \frac{1}{\Gamma_{{\cal R}{\cal R}}^{{\cal R}}} \simeq
2 \frac{\tilde{K}_{{\cal R}\bar{\cal R}}}{\partial_{{\cal R}}\tilde{K}_{{\cal R}\bar{\cal R}}} ~~~
(\nabla_{{\cal R}} G_{{\cal R}} = - G_{{\cal R}\bar{{\cal R}}} ).
\label{SUSYbm} 
\end{eqnarray}
For the second derivatives $\partial_I \partial_J V_F $, we obtain
\begin{eqnarray}
V_{X\bar{X}} &=& e^G (2 - R_{X\bar{X}X\bar{X}}|G^X|^2 )  ,~~~
V_{rr} = 4e^G (2 \tilde{K}_{{\cal R}\bar{{\cal R}}} - R_{{\cal
    R}\bar{{\cal R}}{\cal R}\bar{{\cal R}}}|G^{{\cal R}}|^2),
\nonumber \\
V_{XX} &\sim & e^G , ~~~ V_{r X} = 0.
\end{eqnarray}
Here we have denoted ${\cal R} = r +is$.
When one sets $\hat K$ as (\ref{eq:K-corr}), one obtains
\begin{eqnarray}
- R_{X\bar{X}X\bar{X}} \simeq  \frac{1}{\Lambda^2} \gg 1. 
\end{eqnarray}
With respect to the SUSY breaking ($R$-)axion, let us take the K\"ahler potential below
%assuming that $\tilde{K}_{{\cal R}\bar{{\cal R}}}$ gets singular in ${\cal R} \to 0$ limit (bulk modulus), 
%say 
\begin{eqnarray}
\tilde{K} \equiv -n \log ({\cal R} + \overline{\cal R}) + \delta ({\cal R} + \overline{\cal R}) 
- ({\cal R} + \overline{\cal R}),
\end{eqnarray}
and we write 
\begin{eqnarray}
\tilde{K}_{{\cal R}\bar{{\cal R}}} \equiv 
n \frac{[ 1 + \Delta ({\cal R} + \overline{\cal R})]}{({\cal R} + \overline{\cal R})^2} .
\end{eqnarray}
Then, one can find
\begin{eqnarray}
G^{{\cal R}} &\simeq & -({\cal R} + \overline{\cal R})\left(
1 + \frac{1}{2} ({\cal R} + \overline{\cal R}) \cdot \Delta ' 
\right) , \nonumber \\
- R_{{\cal R}\bar{{\cal R}}{\cal R}\bar{{\cal R}}} 
&\simeq & -2 n \frac{1}{({\cal R} + \overline{\cal R})^4} \left( 
1 + \Delta + \frac{1}{2}({\cal R} + \overline{\cal R})^2 \Delta ''
\right), \\
V_{rr} &\simeq &  - 4 
n \frac{( 2 \Delta ' + ({\cal R} + \overline{\cal R}) \Delta ''  ) }
{({\cal R} + \overline{\cal R})} e^G  \sim 
\Delta \tilde{K}_{{\cal R}\bar{{\cal R}}} e^{G}
.\nonumber 
\end{eqnarray}
Here we used eq.(\ref{SUSYbm}) and $\Delta$ would come from the construction effect of ${\cal R}$ from the original moduli
or the quantum effects of order $g_s$ and of order $\alpha '$
and would be expected as
\begin{eqnarray}
\Delta 
%=\sum_{m > 0}  \frac{a_m}{({\cal R} + \overline{\cal R})^m} 
\lesssim O(1) .
\end{eqnarray}

This result is applicable to many scenarios 
including the SUSY breaking light ($R$-)axion multiplet 
\cite{Balasubramanian:2005zx, Acharya:2006ia, Abe:2005pi, vonGersdorff:2005bf}.
For the above case, fine-tuning of the vanishing cosmological constant 
leads to
\begin{eqnarray}
|G_X|^2 + G_{{\cal R}}G^{{\cal R}} 
\simeq |G_X|^2 + n + O(\Delta) =3 .
\end{eqnarray}
Then one should set
\begin{eqnarray} 
G_X \simeq \sqrt{3 - n +O(\Delta)} ,
\end{eqnarray}
where $n > 0$.
Thus we obtain
\begin{eqnarray} 
F^X \simeq - \sqrt{(3 - n) +O(\Delta)} m_{3/2} ,~~~
\frac{F^{{\cal R}}}{{\cal R}+\overline{\cal R}} \simeq m_{3/2},~~~
\frac{F^i}{\Phi^i + \overline{\Phi^i}} \simeq m_{3/2} \frac{m_{3/2}}{m_{\Phi^i}}.
\end{eqnarray}
For $n=3$, the sGoldstino is almost the SUSY breaking ($R$-)saxion.

Here non-holomorphic sGoldstino mass is given by
\begin{eqnarray}
m^2 &=& 3\hat{\sigma}m_{3/2}^2 , \nonumber \\
\hat{\sigma} &=& \frac{2}{3} - \frac{1}{9} 
\left(
R_{X\bar{X}X\bar{X}}|G^X|^4 + R_{{\cal R}\bar{{\cal R}}{\cal R}\bar{{\cal R}}}|G^{{\cal R}}|^4 
\right) \\
&\sim & \frac{2}{3} + \frac{1}{9\Lambda^2}(3-n)^2 -\frac{2n}{9} + O(\Delta ).
 \nonumber
\end{eqnarray}
Then so long as $V_{rr} > 0$, we would obtain positive definite mass matrix for $n \neq 3$
\begin{eqnarray}
m_{X^{\pm}} \simeq \frac{m_{3/2}^2}{\Lambda^2},~~~
m^2_{r} = \frac{1}{2}\frac{V_{rr}}{\tilde{K}_{{\cal R}\bar{{\cal R}}}} 
\sim \Delta m_{3/2}^2 .
\end{eqnarray}
For $n = 3$, one finds
\begin{eqnarray}
m_{X^{\pm}} \sim m_{3/2}^2 \left( 1 + \frac{\Delta}{\Lambda^2} \right),~~~
m^2_{r} 
\sim \Delta m_{3/2}^2 .
\end{eqnarray}

Here, after the Goldstino is absorbed into the gravitino, the unnormalized axino masses are given by 
\begin{eqnarray}
(m_{\tilde{a}})_{{\cal R}{\cal R}} 
&=& e^{G/2}\left[\nabla_{{\cal R}} G_{{\cal R}} + \frac{1}{3}G_{{\cal
      R}} G_{{\cal R}} \right] \nonumber \\
&\simeq & e^{G/2} \left[- G_{{\cal R} \bar{{\cal R}}} 
+ \frac{1}{3}G_{{\cal R} \bar{{\cal R}}} G^{{\cal R}} G_{{\cal R}} \right] \\
&\simeq & e^{G/2}  G_{{\cal R} \bar{{\cal R}}} \left[- 1
+ \frac{n}{3} + O(\Delta)\right] .\nonumber 
\end{eqnarray}
For $n=3$, SUSY breaking ($R$-)axino becomes the Goldstino, which is absorbed into the gravitino.

We give a comment on the small mixing $G_{{\cal R}i} =G_{{\cal R}\bar{i}} \neq 0$ here.
In many cases, there is the off-diagonal K\"ahler metric $G_{{\cal R}i} = G_{{\cal R}\bar{i}}\neq 0$ and
the main source of the SUSY breaking could be the overall (volume) modulus $(n=3)$
and it affects $F$-term of heavy moduli $\Phi^i$ if any:
$G^i \sim (G^{-1})^{ij}G^{{\cal R}}\nabla_j G_{{\cal R}} \sim (\Phi^i + \overline{\Phi^i})m_{3/2}/m_{\Phi^i}$.
Here we have used the explicit K\"ahler potential for the LARGE volume case in Appendix. 
However, as a consequence, the qualitative features in this section include such scenarios.
Thus the result in this section would be applicable to such cases.

%\subsection{With a constant or $R$-breaking operator}
%
%Suppose that we can assigin the $R$-symmetry except for the constant $W_0$ in the superpotential.
%
%Here if we add the SUSY breaking field $X$ which is a so-called $F$-term uplifting field,
%we find the superpotential
%\begin{eqnarray}
%W &=& W_0 + e^{-R}{\cal W} ,
%\end{eqnarray}
%where $ \hat{W} = e^{a\phi}{\cal W} $ denotes moduli superpotential,
%where only $R$ can has non-linear $R$-charge $2$: $R \to R + i2\alpha$ under the $R$-symmetry.

\section{Corrections to axion masses}

Axions $b^\alpha$ are exactly massless at the previous stage. 
Here, let us consider small corrections to the axion masses.
These can be computed also in the SUSY vacuum,
%since the similar mass correction will be obtained 
if the SUSY breaking sector does not violate any continuous PQ
symmetry of $u^{\alpha}$.
Recall that 
only heavy moduli should be coupled to the SUSY braking sector 
except the $R$-axion.
%and SUSY breaking moduli.
For the small corrections, shifts of the saxion masses are negligible.

\subsection{Superpotential correction}

Here, we consider the correction term $\delta W (\Phi^i,u)$ 
to the previous superpotential (\ref{spotential1}).
That is, we study the following superpotential: 
\begin{eqnarray}
W = {\cal W}(\Phi^i) + \delta W (\Phi^i,u) ,
\end{eqnarray}
where
\begin{eqnarray}
{\cal W}(\Phi^i) &=& W_0+ \sum_k  A_k \exp(-\sum_i a^{(k)}_i \Phi^i)
, 
\nonumber\\
\delta W (\Phi^i,u) &=&  \sum_k  B_k 
\exp(-\sum_{\hat i} b^{(k)}_{\hat  i} \Phi^{\hat i}) .  
\end{eqnarray}
Recall that $\Phi^{\hat i}$ denote all of the moduli including 
$\Phi^i$ and $u^\alpha$.
Hence, the term ${\cal W}(\Phi^i)$ includes only heavy moduli
$\Phi^i$, 
but not light axion multiplets $u^\alpha$, while $\delta W (\Phi^i,u)$ 
includes $u^\alpha$.
We assume $B_k \simeq A_k ={\cal O}(1)$.
We would like to consider the situation that 
$\langle {\cal W} \rangle \gg \langle \delta W \rangle$.
If any terms $B_k \exp(-\sum_{\hat i} b^{(k)}_{\hat i} \Phi^{\hat i})$ 
in $\delta W (\Phi,u)$ do not satisfy the condition, 
$\langle {\cal W} \rangle \gg  B_k \exp(-\sum_{\hat i}
b^{(k)}_{\hat i} \langle \Phi^{\hat i} \rangle ) $,   
we have to take into account such terms from the previous stage 
of moduli stabilization in sections 2 and 3  
and include them in ${\cal W}(\Phi)$.
Then, some of $u$ become heavy reducing the number of light axions.
Therefore heavy moduli should be coupled to saxion-axion multiplets in $\delta W$.
%and it
%would originate also from the higher order effect of instantons or
%Veneziano-Yankielowicz effective action which breaks the $R$-symmetry explicitly,
%so that $\langle {\cal W} \rangle \gg \langle \delta W \rangle$ is obtained
%\footnote{
%We would like to thank O. Lebedev for pointing this out.
%}.

Then one finds the axion mass $m_a$ as \cite{Conlon:2006tq, Acharya:2010zx}
\begin{eqnarray}
{\cal L} &=& 
-K_{\alpha \bar{\beta}} 
\partial_{\mu} b^{\alpha}\partial^{\mu} b^{\beta} - (m_a^2)_{\alpha
  \beta} b^{\alpha} b^{\beta}, \nonumber \\
 (m_{a}^2)_{\alpha \beta} 
&=& 3 e^{K}|W|^2{\rm Re}\left(\frac{\delta W_{\alpha \beta}}{W}\right) ,
\end{eqnarray}
where $\delta W_{\alpha \beta}/W \gg \delta W_{\alpha}\delta W_{\beta}/W^2$ can be obtained in such vacua.

Now we parametrize $\delta W/W$, 
in particular, $b^{(k)}_{\hat i} \langle \Phi^{\hat i} \rangle
\ln \langle {\cal W} \rangle$.
For that purpose, we choose a typical term, say,
$A_j \exp(-\sum_i a^{(j)}_i \Phi^i)$ in ${\cal W}$, which 
represents the value of $\langle {\cal W} \rangle$, i.e.
$A_j \exp(-\sum_i a^{(j)}_i \langle \Phi^i \rangle ) \sim {\cal W}$.
Then, we use the following parameters,
\begin{eqnarray}
r_k = \frac{\sum_{\hat i} b^{(k)}_{\hat i} \langle \Phi^{\hat i} \rangle}{ 
\sum_i a^{(j)}_{i} \langle \Phi^i \rangle }.
\end{eqnarray}
The parameters would satisfy $r_k > 1$, 
because $\langle {\cal W} \rangle \gg \langle \delta W \rangle$.
It is expected that $r_k$ is of ${\cal O}(1)$ or could be a few tens.
Using these parameters, we write 
$B_k \exp(-\sum_{\hat i} b^{(k)}_{\hat i} \Phi^{\hat i})$ in $\delta W$ as 
\begin{eqnarray}
B_k \exp(-\sum_{\hat i} b^{(k)}_{\hat i} \Phi^{\hat i}) 
\simeq W \left( \frac{m_{3/2}}{M_{\rm Pl}} \right)^{r_k - 1}.
\end{eqnarray}
%where $\sum_i a^{(j)}_{i} \Phi^i$ is the exponent in 
%we parametrize $\delta W$ as
%\begin{eqnarray}
%\delta W &=& \sum_k  B_k \exp(-\sum_i b^{(k)}_i \Phi^i) \\
%& \equiv & \sum_k  B_k \exp(- r_k \sum_J a^{(I)}_{J}\Phi^J)  \\
%& = &  \exp(-  \sum_J a^{(I)}_{J}\Phi^J) \sum_k B_k \exp(- (r_k - 1 ) \sum_J a^%{(I)}_{J}\Phi^J)  \\
%& \simeq & W \sum_k B_k \exp(- (r_k -1 )\sum_J a^{(I)}_{J}\Phi^J) \\
%& \simeq & W \sum_k \left( \frac{m_{3/2}}{M_{\rm Pl}} \right)^{r_k - 1} .
%\end{eqnarray}
%Here we assumed in the vacuum
%\begin{eqnarray}
%W &\simeq & W_0 \sim A_I \exp(-\sum_J a^{(I)}_{J}\Phi^I) \gtrsim \sum_{j \neq I}A_j \exp(-\sum_i a^{(j)}_{k}\Phi^k), \\
%\sum_i b^{(k)}_i \Phi^i &\equiv & r_k \sum_J a^{(I)}_{J}\Phi^J .
%\end{eqnarray}
%$r_{k}$ will become important.
Thus, the axion masses with the canonical normalization are given by 
\begin{eqnarray}
(m_a^2)_{\alpha \alpha} \simeq 3 \frac{e^K |W|^2}{K_{\alpha
    \bar{\alpha}}} 
{\rm Re}\left(\frac{\delta W_{\alpha \alpha}}{W}\right)
\simeq 3 \frac{b_{\alpha}^2}{f_{\alpha}^2} m_{3/2}^2 \left( \frac{m_{3/2}}{M_{\rm Pl}} \right)^{r_{\alpha}-1} ,
\end{eqnarray}
if and only if the axion mass is positive definite.
Here we have defined through the diagonalization
\begin{eqnarray}
\delta W_{\alpha \alpha} \equiv b^2_\alpha \delta W ,~~~{K}_{\alpha
  \bar \alpha} \equiv f^2_\alpha ,
\end{eqnarray}
where $f_\alpha =O(M_{\rm string}/M_{\rm Pl})$ are diagonalized decay constants.

%Note that heavy moduli should be coupled to axion multiplets via the superpotential
%to generate smaller scale $\delta W$ than ${\cal W}$ at the same time
%when heavy moduli are stabilized.
%Then the term $\delta W(\Phi, u)$ is replaced by 
%$\delta W(\langle \Phi \rangle , u)$.
%Otherwise, one would not obtain smaller scale: e.g. $b u \sim - \log({\cal W})$.

Once a small value of the gravitino mass $m_{3/2}$ is realized 
such as $m_{3/2} \ll M_{\rm Pl}$, 
the hierarchical axion masses with exponential suppression could appear.
Some examples of mass scales are shown in Table \ref{Tab:axion-mass} 
for $m_{3/2}=1$, $10$ and 100 TeV up to $b_{\alpha}^2/f_{\alpha}^2$.

\begin{table}[ht]
\begin{center}
\begin{tabular}{|c|c|c|c|c|} \hline 
$r_\alpha$ & 3 & 5 & 7 & 9 \\ \hline \hline
$m_a$ for $m_{3/2}=$ 1 TeV & $10^{-4}$ eV & $10^{-19}$ eV & 
$10^{-34}$ eV & $10^{-50}$ eV  \\
$m_a$ for $m_{3/2}=$ 10 TeV & $10^{-2}$ eV & $10^{-16}$ eV & 
$10^{-30}$ eV & $10^{-45}$ eV  \\
$m_a$ for $m_{3/2}=$ 100 TeV & $1$ eV & $10^{-13}$ eV & 
$10^{-26}$ eV & $10^{-40}$ eV  
\\ \hline
\end{tabular}
\caption{Axion masses up to $b_{\alpha}^2/f_{\alpha}^2$. \label{Tab:axion-mass}}
\end{center}
\end{table}

%\begin{table}[ht]
%\begin{center}
%\begin{tabular}{|c|c|c|c|c|} \hline 
%$r_\alpha$ & 4 & 6 & 8 & 10 \\ \hline \hline
%$m_a$ for $m_{3/2}=$ 1 TeV & $10^{-2}$ eV & $10^{-17}$ eV & 
%$10^{-32}$ eV & $10^{-47}$ eV  \\
%$m_a$ for $m_{3/2}=$ 10 TeV & $10^{-1}$ eV & $10^{-15}$ eV & 
%$10^{-29}$ eV & $10^{-43}$ eV  \\
%$m_a$ for $m_{3/2}=$ 100 TeV & $1$ eV & $10^{-13}$ eV & 
%$10^{-26}$ eV & $10^{-39}$ eV  
%\\ \hline
%\end{tabular}
%\caption{Axion masses up to $b_{\alpha}^2/f_{\alpha}^2$. \label{Tab:axion-mass}}
%\end{center}
%\end{table}

We show several illustrating examples to lead to light axion masses 
in what follows.

\begin{itemize}

\item Example 0: $R$-axion mass

The small constant term 
in the superpotential induces the $R$-axion mass\footnote{
There will be also higer order terms from non-perturbative effects breaking the $R$-symmetry,
such like $\omega e^{-2 {\cal R}}$ in the $\delta W$ where $\langle \omega \rangle \sim \langle {\cal W}^2 \rangle$.
But the discussion is similar to the case that $W_0 \sim \langle \omega e^{-2 {\cal R}} \rangle$.
}, 
\begin{eqnarray}
W= {\cal W}e^{-{\cal R}} + W_0 .
\end{eqnarray}
For $W_0 \ll {\cal W}e^{-{\cal R}} \sim Ae^{-a\Phi}$, 
one finds
\begin{eqnarray}
m^2_a &\sim&  e^K \frac{W_0 {\rm Re}(e^{-{\cal R}}{\cal W})}{{\cal K}_{{\cal R}\bar{{\cal R}}}}
\left( \frac{K_{{\cal R}}}{{\cal K}_{{\cal R}\bar{{\cal R}}}} +
  O(1)\right) \nonumber \\
&\sim & \frac{m_{3/2}^2}{{\cal K}_{{\cal R}\overline{{\cal R}}}} 
{\rm Re}\left(\frac{W_0}{e^{-{\cal R}}{\cal W}}\right) 
\left( \frac{K_{{\cal R}}}{{\cal K}_{{\cal R}\bar{{\cal R}}}} + O(1)\right).
\end{eqnarray}
This result also coincides with the result of field-theoretic $R$-axion with 
$e^{-{\cal R}} \equiv \phi$ and $K = \overline{\phi}\phi$
even for larger $W_0 \simeq {\cal W}e^{-{\cal R}}$ .
%
%For $W_0 \ll {\cal W}e^{-{\cal R}}$, one finds
%\begin{eqnarray}
%m^2_a \sim 
%\frac{m_{3/2}^2}{{\cal K}_{{\cal R}\overline{{\cal R}}}} 
%{\rm Re}\left(\frac{W_0}{e^{-{\cal R}}{\cal W}}\right) 
%\left( \frac{K_{{\cal R}}}{{\cal K}_{{\cal R}\bar{{\cal R}}}} + O(1)\right).
%\end{eqnarray}
%This agrees with the above formula via the K\"ahler transformation $\delta W = e^{{\cal R}}W_0$.
On the other hand, 
for $W_0 \gtrsim {\cal W}e^{-{\cal R}}$, one finds the heavy $R$-axion like KKLT
which is stabilized near the SUSY solution
\begin{eqnarray}
m^2_a \sim 
{m_{3/2}^2} 
\left(\frac{K_{{\cal R}}}{{\cal K}_{{\cal R}\bar{{\cal R}}}}\right)
\left( \frac{K_{{\cal R}}}{{\cal K}_{{\cal R}\bar{{\cal R}}}} + O(1)\right) .
\end{eqnarray}
Here one finds ${\cal W}e^{-{\cal R}} \sim {\cal K}_{{\cal R}} W_0 $ for the KKLT stabilization.

\item Example 1: $SU(N+M) \times SU(M)$ gaugino condensations or with an instanton ($M=1$)

Let us consider the KKLT type superpotential \cite{Conlon:2006tq, Bobkov:2010rf}:
\begin{eqnarray}
W = W_0 + e^{-a\Phi} + e^{-b(u+\Phi)},~~~
a= \frac{8\pi^2}{N+M} ,~~~b= \frac{8\pi^2}{M} ,~~~N \gg M, 
\end{eqnarray}
where $\Phi$ is the heavy modulus and $u$ is the light saxion-axion multiplet.
In this case, assuming $\langle u \rangle \lesssim \langle \Phi \rangle$,
one obtains $r \sim N/M + 1$ and the axion mass is estimated as 
\begin{eqnarray}
m^2_a \sim 3 \frac{b^2}{f^2} m_{3/2}^2 \left( \frac{m_{3/2}}{M_{\rm Pl}} \right)^{\frac{N}{M}} .
\end{eqnarray}
A similar result can be obtained in the racetrack model
\cite{Acharya:2010zx},
\begin{eqnarray}
W &=& W_0 + e^{-a_1 \Phi} -  e^{-a_2 \Phi} + e^{-b(u+\Phi)}, \\
a_{1,2} &=& \frac{8\pi^2}{N_{1,2}+M} ,~~~b= \frac{8\pi^2}{M} ,~~~
N_1 \sim N_2 ,~~~ N_{1,2} \gg M ,
\end{eqnarray}
when we do not fine-tune $W_0$ as a special value.

\item Example 2: Many gaugino condensations or instantons wrapping on multiple cycles (in intersecting D-brane system)

Consider the superpotential with $n+3$ moduli; one is heavy modulus
$\Phi$ and the remaining $n+2$ multiplets include light axions $u_I$
\begin{eqnarray}
W &=& W_0 + e^{-a\Phi} + \sum_{i=1}^{n+2} \exp[-b_i (\sum_{I \neq i}^{n+2} u_I) -b\Phi],\\
&& a \sim b_i ~~~{\rm for}~^\forall ~i, ~~~n \gg 1.
\end{eqnarray}
In this case, if $\langle u_i \rangle \sim \langle \Phi \rangle~{\rm
for}~^\forall ~i~$, one finds $r \sim n+1$ and the axion mass 
is estimated as 
\begin{eqnarray}
m^2_a \sim 3 \frac{b^2}{f^2} m_{3/2}^2 \left( \frac{m_{3/2}}{M_{\rm Pl}} \right)^{n} .
\end{eqnarray}
However, if $\langle u_i \rangle \ll \langle \Phi \rangle~{\rm for}~^\forall ~i~$,
one cannot obtain small axion masses; one needs $a \ll b$ as the previous example.

\item Example 3: Including gaugino condensation on the magnetized brane

One may obtain the superpotential on the magnetized D7-branes or E3-branes 
wrapping on the divisor $D$
in type IIB orientifold:
\begin{eqnarray}
W &=& W_0 + e^{-a\Phi} + \hat{B} e^{-b(u+\Phi)},~~~ \hat{B} = B\exp[- b {\cal M} \langle S \rangle] .
\end{eqnarray}
Here the constant ${\cal M}$ denotes ${\cal M} = \frac{1}{8\pi^2}\int_D {\cal F}^2 \in \mathbb{Z}$
up to curvature term \cite{Haack:2006cy},
%which is proportional to $\chi(D)/24$, 
%where $D$ is the divisor on which a stack of the D7-branes is wrapping,
%and $\chi$ is the Euler characteristics of $D$
${\cal F}$ is the world volume flux and $\langle S \rangle $ is the vev of the complex dilaton, which is fixed by three form flux.
In this case, if $b \langle S \rangle \sim b\langle u \rangle \sim
a\langle \Phi \rangle$, one can find $r \sim {\cal M} + 1$ and 
the axion mass is estimated as 
\begin{eqnarray}
m^2_a \sim 3 \frac{b^2}{f^2} m_{3/2}^2 \left( \frac{m_{3/2}}{M_{\rm Pl}} \right)^{{\cal M}} .
\end{eqnarray}
A value of ${\cal M}$ is weakly constrained via the tadpole condition of D3-branes in the F-theory limit 
of the orientifold compactification \cite{Blumenhagen:2008zz}:
\begin{eqnarray}
N_{D3} + \frac{1}{2}N_{\rm flux}({\cal M}) = \frac{\chi(Y_4)}{24}.
\end{eqnarray}
Here 
%r.h.s. can be typically of $O(1000)$ and 
$Y_4$ is an elliptically fibred Calabi-Yau four-fold.
On the other hand, it would be natural and more plausible that $u \gtrsim {\cal M} S$ on the D7-brane, i.e. ${\cal M}=O(1)$. 
However, with the T-dual description, the present case would also be plausible since ${\cal M}$ corresponds to a winding number.

\end{itemize}

Thus, many models could lead to the hierarchical axion masses 
with suppression, $r={\cal O}(1)$ or a few tens.

\subsection{ K\"ahler potential correction}

Here, we comment on corrections to axion masses from the K\"ahler potential.
Suppose that
\begin{eqnarray}
K = {\cal K}(\Phi + \overline{\Phi}) + \delta K (\Phi,
\overline{\Phi}) ,
\end{eqnarray}
where $\delta K (\Phi,\overline{\Phi}) $ is a correction term and 
%\begin{eqnarray}
$
\frac{\partial \delta K}{\partial b^{\alpha}} \neq 0.
$
%\end{eqnarray}
Then one finds the axion masses \cite{Conlon:2006tq}
\begin{eqnarray}
 (m_{a}^2)_{\alpha \beta} 
&=& 3 e^{K}|W|^2 \left[
-\delta K_{\alpha \bar{\beta}}+
{\rm Re}(\delta K_{\alpha \beta}) 
\right].
\end{eqnarray}
Here we have neglected $O((\delta K)^2)$ term in the vacuum.
%{\bf In this paper we are neglecting smaller mass scale than the gravitino mass.
%Then even if there is a correction $\delta K$, 
%one can neglect this contribution for $\delta K'' /K'' < 1 $.
%our computation is plausible when $\delta K'' /K'' < 1 $.
%Note that one can also neglect the corrections
%compared to those from the superpotential, for $\delta K''  \lesssim
%\delta W '' / W $.
%} 
It is plausible that $\delta K (\Phi,\overline{\Phi}) $ would also 
appear from non-perturbative effects such as 
\begin{eqnarray}
\delta K (\Phi,\overline{\Phi}) = \sum_k  B'_k 
\exp(-\sum_{\hat i} b'^{(k)}_{\hat  i} \Phi^{\hat i}) + h.c.  
\end{eqnarray}
In this case, the hierarchical axion masses with 
exponential suppression would be obtained similarly to 
the superpotential corrections.

\section{Comment on axions from matter-like fields}

Here, we (briefly) study the superpotential including matter-like fields below:
\begin{eqnarray}
W &=&  {\cal W} + \delta W, 
\end{eqnarray}
where
\begin{eqnarray}
{\cal W } \equiv  W_0 + \sum_k  A_k ({\Psi}) \exp(-\sum_i a^{(k)}_i
\Phi^i), \nonumber \\
\delta W \equiv \sum_k  B_k ({\Psi}) \exp(-\sum_i b^{(k)}_i \Phi^i) .
%\\
%&\equiv & {\cal W} + \Delta W  .
\end{eqnarray}
Here we omitted SUSY breaking sector and
$\{ {\Psi} \}$ means matter fields (or open string moduli) originating from the open string.
We assume that the matter fields stabilized near the SUSY solution $K_{\Psi} W \sim W_{\Psi}$. 
%
%More precisely, as the matter K\"ahler potential
%is typically given by $K \propto |\Psi|^2$, one can expect
%\begin{eqnarray}
%F^{\Psi} \sim e^{G/2} \left( K_{\bar\Psi} + \frac{\overline{W_{\Psi}}}{\overline{W}}\right) \sim m_{3/2} \Psi . 
%\end{eqnarray}
%when the off-diagonal K\"ahler metric, e.g. $K_{\Psi \bar{\Phi}}$, is negligible. 
%
Let us focus on the light matter-like fields whose 
axionic parts are massless while saxions are stabilized
%assuming that the heavy fields and saxion fields are stabilized 
e.g. via $F$-term, $D$-term conditions
or quantum radiative corrections.
At low energy they are written by 
\begin{eqnarray}
{\Psi}^P &\equiv &  |\langle \Psi^P \rangle| e^{-\psi^P} 
%,~~~~{\rm where}~~\psi^P = \varphi^P + i \vartheta^P
, \nonumber \\
%K &\sim & e^{-(\psi + \bar{\psi})},~~~
A_k(\Psi) &=& \prod_P |\langle \Psi^P \rangle| e^{-n_P^k \psi^P}, ~~~
B_k(\Psi) = \prod_P |\langle \Psi^P \rangle| e^{-m_P^k \psi^P} .
\end{eqnarray}
Here some $A_k$ or $B_k$ can be constants.
Consider linear combinations of $\hat{\psi}^p = \sum_P c_{~P}^p \psi^P$, 
i.e. ${\psi}^P = \sum_p (c^{-1})^{P}_{~p} \hat{\psi}^p$ \footnote{
One can consider a linear combination including closed string moduli
when matter-like fields are coupled to light moduli via a non-perturbative effect.
For simplicity we will not consider such a case.
}, 
such that one can find
\begin{eqnarray}
\frac{\partial {\cal W}}{\partial \hat{\psi}^p} =0, \qquad 
 \frac{\partial A_k (\Psi)}{\partial \hat{\psi}^p} = 0~~~{\rm for}~~^\forall k,~^\exists p.
\end{eqnarray}
We use the K\"ahler potential at the tree level as
\begin{eqnarray}
K &=& \sum_P Z_P (\Phi + \bar\Phi) \overline{\Psi^P} \Psi^P .
\end{eqnarray}
Then one finds at low energy \cite{Higaki:2011bz}
\begin{eqnarray}
K & = &\sum_P Z_P (\Phi + \bar\Phi) |\langle \Psi^P \rangle|^2
e^{-(\psi^P +\overline{\psi^P})}\nonumber  \\
&\equiv & 
\lambda_p (\hat{\psi}^p +\overline{\hat{\psi}^p}) 
+ \frac{\lambda_{pq}}{2}(\hat{\psi}^p +\overline{\hat{\psi}^p}) (\hat{\psi}^q +\overline{\hat{\psi}^q})
+ O((\hat{\psi} +\overline{\hat{\psi}})^3),
\end{eqnarray}
where
\begin{eqnarray}
\lambda_p = - \sum_P (c^{-1})^{P}_{~p}|\langle \Psi^P \rangle|^2 ,~~~
\lambda_{pq} = \sum_P (c^{-1})^{P}_{~p} (c^{-1})^{P}_{~q} |\langle \Psi^P \rangle|^2 .
\end{eqnarray}
For simplicity, we set $\lambda_p,~\lambda_{pq} = {\rm const.}$ in the vacuum, i.e. 
$Z_P ={\rm const.}$ and would depend on much heavier moduli vevs.
If $Z_P$ depends on the moduli or there are mixings between $\psi$ and $\Phi$ in $\hat\psi$, 
$\lambda_p$ and $\lambda_{pq}$ also have the dependence on the closed string moduli.

Then for axion multiplets one finds\footnote{
For light non-axion multiplets which have of $O(m_{3/2})$ masses, they can be
stabilized through the superpotential ($\partial_q {\cal W} \neq 0$) and
have the similar properties to those of axion multiplets \cite{Choi:2011xt}.
For instance one can obtain
$G_q = K_q +\frac{W_q}{W} \sim \lambda_q$.
Here $W_q \sim K_q W$.
Therefore $\hat{\psi}^p$s can include such light non-axion modes.
}
\begin{eqnarray}
G_p = K_p = \lambda_p .
\end{eqnarray}
Thus even if $G_p = \lambda_p = 0 $ for axion multiplets $\hat{\psi}^p$,
because of $\lambda_{pq} \neq 0$ one finds
\begin{eqnarray}
F^p = e^{G/2} K^{p\overline{q}}G_{\overline{q}}
= e^{G/2}(\lambda^{-1})^{pq}\lambda_q
\sim m_{3/2} ,
\end{eqnarray}
which leads to 
${F^{\Psi}}/{\Psi} \sim  m_{3/2} $.
This result is consistent with the stationary condition $\partial_p V = 0$:
$G_p + G^q \nabla_p G_q = 0$, which leads to  $G^q  = O(1)$.

Then the saxion masses can be found 
\begin{eqnarray}
%\partial_{\varphi^p} V_F &=& 2G_{p}V_F + 2e^G [2 G_{p} + G_q G_{\bar{r}} \partial_p G^{q\bar{r}}] \\
\langle \partial_{\varphi^p} \partial_{\varphi^q} 
V_F \rangle &=&
4 e^{G} \left[
2 G_{p\bar{q}} - \partial_{r}G_{p\bar{q}}G^r 
- \bar{\partial}_{\bar{r}}G_{p \bar{q}}G^{\bar{r}} + G_r G_{\bar{s}}
\partial_{p} \bar{\partial}_{\bar{q}}G^{r\bar{s}}
\right] \sim e^{G}G_{p \bar{q}}
,
\end{eqnarray}
where $\hat{\psi}^p = \varphi^p + i \vartheta^p$.
Whether $\langle \partial_{\varphi^p} \partial_{\varphi^q} V_F \rangle >0$ or $<0$ 
depends on the model, but typical order of the saxion masses are of $O(m_{3/2})$ even though
the masses can also receive the contribution from D-terms of anomalous $U(1)$ symmetries \cite{LARGE2-2}.
When there is the vanishing saxion mass at the tree level,
quantum radiative correction induces the mass smaller than $m_{3/2}$ 
\cite{Murayama:1992dj, Nakamura:2008ey, Choi:2011xt}.
Note that from the assumption that $Z_P ={\rm const.}$, 
one finds $V_{\varphi^p X} = V_{\varphi^p i} =V_{\varphi^p \alpha} =0$.

The axion masses induced by $\delta W$ depend on the model, i.e. 
vevs of closed string moduli, those of matter like fields or 
the power of polynomial of matter-like fields in the superpotential.
After the Goldstino is absorbed into the gravitino, the unnormalized axino masses are given by 
\begin{eqnarray}
(m_{\tilde{a}})_{pq} 
&=& e^{G/2}\left[\nabla_{p} G_{q} + \frac{1}{3}G_{p} G_{q} \right]
\nonumber \\
& \sim & e^{G/2} G_{p \bar{q}} .
\end{eqnarray}

%%%%%%%%%%%%%%%%%%%%%%%%%%%%%%%%%%%%%%%%%%%%%%%%%%%%%%%%%%%%%%%%%%%%%%%%%%%%

%%%%%%%%%%%%%%%%%%%%%%%%%%%%%%%%%%%%%%%%%%%%%%%%%%%%%%%%%%%%%%%%%%%%%%%%%%%%
%%%%%%%%%%%%%%%%%%%%%%%%%%%%%%%%%%%%%%%%%%%%%%%%%%%%%%%%%%%%%%%%%%%%%%%%%%%%

\section{Conclusion and discussion}

We have studied properties of low energy moduli
stabilization in the ${\cal N}=1$ effective SUGRA, which have 
heavy moduli and would-be saxion-axion multiplets.
We have given general formulation for the scenario, 
where heavy moduli and saxions are stabilized and 
axions remain light.
SUSY breaking effects are important.
In the non-supersymmetric Minkowski vacuum, the stable vacuum can be obtained 
even though there are light string-theoretic axions.
In such a vacuum, heavy moduli and saxions can be stabilized
supersymmetrically.
In particular, saxions can be stabilized at the point $K_\alpha \sim
0$, while axions in the same multiplets remain lighter than 
the gravitino mass $m_{3/2}$.
This scenario predicts the same number of saxions with the mass 
$2m_{3/2}$ as the number of light axions.
Note that our analysis on moduli stabilization 
is applicable even if there are not light axions in the vacuum.

When there are some moduli mixing the SUSY breaking source 
in the superpotential, such moduli would also 
destabilize the vacuum.
%contribute SUSY breaking.
In order to avoid such a situation, we need quite heavy masses 
for moduli.
The moduli masses, which are generated in the KKLT-like model, 
are not enough, but one needs heavier masses, which 
would be generated through the racetrack model, D-term or closed string fluxes.

Alternatively, some moduli may contribute to SUSY breaking, 
e.g. the $R$-axion multiplet.
In this case, the saxion mass 
can be lighter than the gravitino.

We have studied the effective SUGRA theory to lead to 
the axiverse.
Following our realization, it is important to study 
further cosmological and particle phenomenological implications.
In addition, our scenario predicts the same number of 
saxions with the mass $2m_{3/2}$ as the number of light axions.
These saxions would also have important implications 
depending on their masses, $2m_{3/2}$.
For example, when $m_{3/2}$ is around ${\cal O}(1)-{\cal O}(100)$ TeV, 
the late time entropy production by the vast number ($\sim 100$) of saxion decays 
into radiations much before the BBN epoch can dilute harmful gravitino abundance 
\cite{Acharya:2006ia} produced by decays of scalar fields such as heavy moduli \cite{Endo:2006zj}.
(See \cite{Lebedev:2006qq, DineKitanoetal, Lebedev:2006qc}
for discussions of the dilution by the SUSY breaking field $X$, which does not decay into gravitinos, 
based on the KKLT stabilization and
see also \cite{Moroi:1999zb, Nakamura:2007wr} for the relevant discussions.)
It is interesting to study other aspects of axions and/or saxions 
following our realization of the axiverse.

We have discussed general aspects of low-energy effective SUGRA 
theory without fixing explicit string models.
It is important to study explicit string model building 
leading to our scenario 
with moduli stabilization and light axions.
We would study explicitly such string models elsewhere.

{\bf Acknowledgement}

The authors would like to thank H.Kodama for useful discussions.
The authors would like also to thank O. Lebedev for reading this maniscript and useful comments.
T.~K. is supported in part by the Grant-in-Aid for 
Scientific Research No.~20540266 and the Grant-in-Aid for the Global COE 
Program "The Next Generation of Physics, Spun from Universality and 
Emergence" from the Ministry of Education, Culture,Sports, Science and 
Technology of Japan.
This work is supported by  the JSPS Grant-in-Aid for Scientific Research
(A) No. 22244030.

%%%%%%%%%%%%%%%%%%%%%%%%%%%%%%%%%%%%%%%%%%%%%%%%%%%%%%%%%%%%%%%%%%%%%%%%%%%%
%%%%%%%%%%%%%%%%%%%%%%%%%%%%%%%%%%%%%%%%%%%%%%%%%%%%%%%%%%%%%%%%%%%%%%%%%%%%
%%%%%%%%%%%%%%%%%%%%%%%%%%%%%%%%%%%%%%%%%%%%%%%%%%%%%%%%%%%%%%%%%%%%%%%%%%%%

\appendix
%{\Large \bf Appendix}

\section{Moduli stabilization models}

Here we review several moduli stabilization models 
in type IIB Calabi-Yau O3/O7 orientifold models.

\subsection{$D$-term stabilization}

We show the relevant part of the model of the $D$-term stabilization.
This is the model with the anomalous $U(1)$ gauge symmetry and
e.g. the blowing-up mode \cite{Higaki:2003zk}
%\footnote{
%We would like to take just $\partial_i K =0$ to have the vanishing Fayet-Iliopoulos term.
%}
\begin{eqnarray}
K &=& \frac{1}{2{\cal V}_E}(M+\overline{M}+V)^2 , ~~~ \partial_M W =0 ,
\end{eqnarray}
where $V$ is the anomalous $U(1)$ vector multiplet and
${\cal V}_E$ is the compactification volume in the Einstein frame: $6 {\cal V}_E = \int J \wedge J \wedge J$, 
where $J$ is the K\"ahler form in the Einstein frame on the Calabi-Yau three-fold.
One can ignore matter-like fields, depending on the charge signature of matter.
Then one finds the minimum via SUSY condition $D_M W = D = K_M = (M+\overline{M})/{\cal V}_E = 0$
and obtains the massive vector multiplet $\tilde{V} =
(M+\overline{M}+V)$, where 
$M$ is eaten by the gauge multiplet.
The mass of the vector multiplet is now given by $g {\cal V}_E^{-1/2}$, where $g$ is the gauge coupling.

\subsection{KKLT}
\label{app:KKLT}

We show the so-called KKLT model \cite{Kachru:2003aw, Conlon:2006tq} with
\begin{eqnarray}
K = -2 \log\left({\cal V}_E \right) ,  \qquad 
W = W_0 + \sum_i^{h^{1,1}_{+}} A_i e^{-a_i T^i} ,
\end{eqnarray}
where $W_0 \ll 1 $.
Here, $h^{1,1}_{+}$ ($h^{1,1}_{-}$) denotes the Hodge number 
of even (odd) parity moduli.
To realize the SUSY breaking Minkowski vacuum, we add 
the uplifting potential,
\begin{eqnarray}
V_{\rm lift} &=& \frac{\epsilon}{{\cal V}_E^{4/3}} , \qquad 
\epsilon \simeq 3   \frac{|W_0|^2}{\langle {\cal V}_E^{2/3}\rangle}  .
\end{eqnarray}
In this case, one finds
\begin{eqnarray}
T_i &\simeq & \frac{-1}{a_i} \log(W_0) \simeq \frac{1}{a_i}
\log\left(\frac{M_{\rm Pl}}{m_{3/2}}\right) , \nonumber \\
\frac{F^{T_i}}{T_i+\overline{T_i}} &\simeq & \frac{m_{3/2}}{a_i\tau_i} \simeq \frac{m_{3/2}}{\log\left({M_{\rm Pl}}/{m_{3/2}}\right)},
\end{eqnarray}
where $\tau_i = {\rm Re}(T_i)$.
The gravitino mass and moduli masses are obtained as 
\begin{eqnarray}
m_{3/2} \simeq \frac{W_0}{{\cal V}_E} , \qquad 
m_{i} \simeq 2 a_i \tau_i m_{3/2} .
\end{eqnarray}

For one bulk volume modulus, we have
\begin{eqnarray}
K &=& -3 \log(T+\overline{T}) , \qquad 
W = W_0 + e^{-a T} , \nonumber \\
V_{\rm lift} &=& \frac{\epsilon}{(T+\overline{T})^2} 
,\qquad 
\epsilon \simeq 3   \frac{|W_0|^2}{\langle (T+\overline{T})\rangle}.
\end{eqnarray}
In this model, anomaly mediation is comparable to $F^T/(T+\overline{T})$.
See also for generalization of this scenario \cite{Abe:2005rx}.

\subsection{LARGE volume scenario}

This is the model \cite{Balasubramanian:2005zx} with bulk moduli and
blowing-up modes, whose K\"ahler potential is written as 
\begin{eqnarray}
K = -\log(S+\overline{S}) -2 \log\left({\cal V}_E +
  \frac{\hat{\xi}}{2}\right) ,
\end{eqnarray}
with
\begin{eqnarray}
{\cal V}_E = (2\tau_b)^{3/2} - \sum_i^{h^{1,1}_+ - 1 } (2\tau_{s,i})^{3/2}.
\end{eqnarray}
Here, for simplicity we have neglected moduli redefinitions at 1-loop level.
The superpotential is written by
\begin{eqnarray}
W = W_0 +\sum_i^{h^{1,1}_{+}-1} A_i e^{-a_i T_s^i} ,
\end{eqnarray}
where $W_0 =O(1)$, and the uplifting potential 
is added as,
\begin{eqnarray}
V_{\rm lift} = \frac{\epsilon}{{\cal V}_E^{4/3}} , \qquad 
\epsilon \simeq  \frac{|W_0|^2}{8 \langle \log({\cal V}_E) {\cal V}_E^{5/3}\rangle} .
\end{eqnarray}
Note that $1/(2\pi g_s) = {\rm Re}(S)$.
We can consider vanishing Standard Model (SM) cycle moduli
or odd parity moduli with $D$-term stabilization: $K = \frac{1}{2{\cal V}_E}(T+\overline{T})^2$. One can consider
the $K3$ fibration model: ${\cal V}_E = \tau_{b,1}\tau_{b,2}^{1/2} - \sum_i^{h^{1,1}_+ - 2 } \tau_{s,i}^{3/2}$ 
together with loop corrections to fix bulk moduli.

Let us consider the simplest case, $h^{1,1} = h^{1,1}_{+}=2$.
In this case, we have
\begin{eqnarray}
{\cal V}_E &\simeq & \tau_b^{3/2} \sim e^{a_s \tau_{s}},~~~
\tau_s^{3/2} \simeq  \hat{\xi} , \nonumber \\
\frac{F^{T_b}}{T_b + \overline{T_b}} &\simeq & m_{3/2} ,~~~
\frac{F^{T_s}}{T_s+\overline{T_s}} \simeq \frac{m_{3/2}}{\log\left({M_{\rm Pl}}/{m_{3/2}}\right)} .
\end{eqnarray}
Here, one finds 
\begin{eqnarray}
m_{3/2} \simeq \frac{W_0}{{\cal V}_E} , \qquad 
m_{b} \sim m_{3/2} \left(\frac{m_{3/2}}{M_{\rm Pl}}\right)^{1/2} ,
\qquad 
m_{s} \sim \log({\cal V}_E) m_{3/2} ,
\end{eqnarray}
with ${\cal V}_E \gg 1$.
Note $a_s\tau_s \sim \log({\cal V}_E) \sim \log\left({M_{\rm Pl}}/{m_{3/2}}\right)$ and
$T_b$ is the SUSY breaking saxion similar to the case \cite{vonGersdorff:2005bf},
whereas the axion is not couple to the visible sector.
$T_b$ is an almost no-scale model modulus, while $T_s$ is a KKLT like modulus.
In this model, anomaly mediation could be suppressed compared to $F^T/(T+\overline{T})$ by ${\cal V}_E^{-r}$, where
$r$ is a fractional number.

\subsubsection{Modified original LARGE volume scenario}

Note that one can consider the model 
such like the original scenario with an additional odd parity moduli 
instead of vanishing the SM cycle on the D7-branes, i.e. 
$h^{1,1}_+ = 2$ and $h^{1,1}_- =1$ and we will discuss 
the neutral stringy instanton or gaugino condensation under the anomalous $U(1)$ symmetries 
on the brane with world volume flux \cite{Grimm:2011dj}.
This is the case in contrast to the paper
\cite{Blumenhagen:2007sm} 
and is similar to the heterotic case \cite{Heteto}.
Then, the K\"ahler potential and the superpotential are 
written by 
\begin{eqnarray}
K = -\log(S+\overline{S}) -2 \log\left({\cal V}_E +
  \frac{\hat{\xi}}{2}\right) , \qquad 
W = W_0 + Ae^{-a(T_+ + qG +hS)}, 
\end{eqnarray}
where 
\begin{eqnarray}
{\cal V}_E = (2\tau_b)^{3/2} - \left(2\tau_{+} +
  \frac{(G+\bar{G})^2}{(S+\bar{S})}\right)^{3/2}.
\end{eqnarray}
Here $G$ is the odd parity K\"ahler moduli
and note that in general odd parity moduli $\{G \}$ necessarily follow even parity moduli $\{T \}$ in the world volume of the brane.
Then we took only the leading term of summation of instanton configuration for simplicity.
In addition, the gauge kinetic function of the SM sector is 
written by 
\begin{eqnarray}
f_{SM} &=& T_+ + q_{SM}G + h_{SM}S .
\end{eqnarray}
Again we neglect moduli redefinitions at 1-loop level.
Here we assume that non-perturbative superpotential comes from the E3-brane instanton wrapping on the divisor $D_E$ with the flux.
$h,~q, ~h_{SM}$ and $q_{SM}$ depend on the flux on E3-brane and the visible sector D7-branes wrapping on $D_{SM}$
holding not only the SM gauge group but also the anomalous $U(1)$ symmetry respectively.
$D_E$ and $D_{SM}$
map to $D_{E'}$ and $D_{SM'}$ respectively under orientifold action;
$D_E$ and $D_{SM}$ include both even and odd elements, e.g. 
$[D^+_{E, SM}] = [D_{E,SM}] + [D_{E',SM'}]$ and $[D^-_{E, SM}] = [D_{E,SM}] - [D_{E',SM'}]$, where $[D]$ is the Poincare dual of $D$. 
Here we take triple intersection $d_{bbb} = d_{+++} = d_{+--} =1$ for simplicity.
Now the presence of $G$ means there can be an anomalous $U(1)$ symmetry;
both $T_{+}$ and $G$ should be charged under the anomalous $U(1)$ symmetry:
\begin{eqnarray}
\delta G = iQ_{G} = i\frac{N}{8\pi^2} ,~~~\delta T_+ = iQ_{T} = -
i\frac{N}{8\pi^2} ({F}^{-}_{D_{SM}} + {\cal F}^+_{D_{SM}} ) .
\end{eqnarray}
Here $N$ is the number of the D7-branes and
$F = F^{-}_{D_{SM}}\omega_- + F^+_{D_{SM}} \omega_+ $ is the internal world volume flux relevant to the anomalous $U(1)$
on the visible sector D7-branes, where ${\cal F}^+ = b^+ + F^+ $ and $b^{+} = 0$ or $1/2$
and $\omega_- \in H^{1,1}_-(CY),~\omega_+ \in H^{1,1}_+(CY)$ are (pull-back on the SM cycle of) 
the harmonic two-cycle basis on the CY space\footnote{
$b^+ =1/2$ would be necessary because of the Freed-Witten anomaly \cite{Freed:1999vc} on the D7-branes wrapping on the $D_{SM}$ 
and the E3-brane.
}. 
Here we took all the wrapping number of the D7-brane and E3-brane against the even or odd cycle unity: 
$C_E^+ = C_E^- =C_{D_{SM}}^+ = C_{D_{SM}}^- =1$ in the notation of the paper \cite{Grimm:2011dj}.
Therefore, the following condition, 
\begin{eqnarray}
q = {F}^{-}_{D_{SM}} + {\cal F}^+_{D_{SM}} ,
\end{eqnarray}
should be satisfied for the neutral superpotential in this simple case\footnote{
Here $f_{SM}$ could be also gauge invariant under the $U(1)$ since we could have $q = q_{SM}$; the $U(1)$ could be non-anomalous. 
However, for instance, when there is a relation that $C^-_{D_{SM}} \neq C^{+}_{E}$, 
or are fluxes depending on the SM gauge group and the $U(1)$,
$f_{SM}$ is not necessarily invariant under the $U(1)$: $q \neq q_{SM}$ and the $U(1)$ is generally anomalous.
}.
The $D$-term potential is given by
\begin{eqnarray}
V_D = \frac{1}{2{\rm Re}(f)}D_A^2, \qquad 
 D_A =  Q_T \partial_T K + Q_G \partial_G K  =
\frac{N}{8\pi^2} (- q \partial_T K + \partial_G K)  ,
\end{eqnarray}
up to matter-like fields.
If all the gauge couplings including the $U(1)$ symmetry are gauge invariant as the above simple case, 
the $U(1)$ can become non-anomalous; one should include matter.
Otherwise, the $U(1)$ is in general anomalous; one would be able to neglect matter.
For such a case, this model would have string theoretic axion, which is absorbed into the $U(1)$ vector multiplet.
Define $\Phi \equiv  T_+ +q G$ and $u \equiv q T_+ -G$.
As a consequence $\Phi$ and $u$ are stabilized near SUSY solution without matter field vevs,
%as the usual case 
$D_{\Phi} W \sim 0$ and $D_A \propto K_{u} \sim 0$;
one obtains the scalar potential after integrating out $u$ and ${\rm Im}(\Phi)$:
\begin{eqnarray}
V \simeq 
\frac{2\sqrt{2}\sqrt{\phi} a^2 \hat A^2 e^{-2 a {\rm Re}(\Phi) } }{3 {\cal V}_E}
- \frac{4\phi a \hat A e^{-a {\rm Re}(\Phi) }W_0}{{\cal V}_E^2}
+ \frac{3 W_0^2 \hat{\xi} }{2 {\cal V}_E^3} ,
\end{eqnarray}
where $\hat A \equiv Ae^{-ahS}$.
Here we have defined $\phi \equiv {\rm Re}(\Phi) - q^2/8$.
Thus one would find $m_{3/2} \sim \frac{W_0}{{\cal V}_E}$ and
\begin{eqnarray}
\nonumber
\langle {\cal V}_E \rangle &\sim & \langle \tau_b^{3/2} \rangle \sim  e^{a(\Phi + hS) } ,~~~
\langle \Phi \rangle \sim  \frac{\hat{\xi}^{2/3}}{2} + \frac{q^2}{8} + i\frac{\pi}{a},~~~ 
\langle {\rm Re}(u) \rangle \sim  q \langle {\rm Re}(\Phi) \rangle  - \frac{1}{4}(q^3+ q ) , 
\\
\nonumber
m_{b} &\sim & m_{3/2} \left( \frac{m_{3/2}}{M_{\rm Pl}} \right)^{1/2} ,~~~
m_{\Phi} \sim  a \phi m_{3/2} \sim \log({\cal V}_E) m_{3/2} ,~~~ m_{u} = M_V \sim Q\frac{M_{\rm Pl}}{\sqrt{{\cal V}_E}} , \\
\frac{F^{b}}{T_b +\overline{T_b}} &\sim & m_{3/2} ,~~~
\frac{F^{\Phi}}{2\phi} \sim  q^{-1}\frac{F^{u}}{2\phi} 
\sim m_{3/2} \frac{m_{3/2}}{m_{\Phi}} \sim \frac{m_{3/2}}{\log({\cal V}_E)} ,~~~
D_A \sim 0 .
%\lesssim O\left( \frac{m_{3/2}^2}{Q{\cal V}_E} \right).
\end{eqnarray}
Thus we find $F^T \simeq F^\Phi$ and $F^G \simeq 0$.
Here we have  used $D_A \sim (\partial_I \bar{\partial}_{\bar{J}}D) F^I \bar{F}^{\bar{J}}/M_{V}^2$ 
\cite{Kawamura:1996wn, Scrucca:2007pj,Choi:2006bh, Choi:2011xt} and
assumed that the anomalous $U(1)$ gauge coupling and vev of the $S$ are of $O(1)$.
One finds in the vacuum
$G^{u}  \simeq qG^{\Phi}$,
$\partial_{T_b}\bar{\partial}_{\bar{\Phi}}K_{u} \simeq -q \partial_{T_b}\bar{\partial}_{\bar{u}}K_{u}$
and $\partial_{\Phi}\bar{\partial}_{\bar{\Phi}}K_{u} \simeq -2q \partial_{\Phi}\bar{\partial}_{\bar{u}}K_{u}$.
Note also that
$\partial_{T_b}\bar{\partial}_{\bar{T}_b}K_u$ and $\partial_{u}\bar{\partial}_{\bar{u}}K_{u}$
are irrelevant\footnote{
Suppose that $D_A \sim K_u \lesssim {\cal V}_E^{-(1+1)} \ll {\cal V}_E^{-1}$. 
Then one can see
$\partial_{T_b}\bar{\partial}_{\bar{T}_b}K_u \simeq K_u/{\cal V}_{E}^{4/3} \lesssim {\cal V}_{E}^{-(7/3+1)}$
and $\partial_{u}\bar{\partial}_{\bar{u}}K_{u} \sim K_u \lesssim {\cal V}_E^{-(1+1)}$ and they are negligible.} 
since one can obtain $K_u \sim 0$ in the vacuum; 
there is be a cancellation in the $D$-term at of $O({\cal V}_E^{-2})$ at least.
Detailed study of this model is beyond the scope of this paper and we will leave it future work.

\subsection{Racetrack model}
\label{app:racetrack}

This is the model \cite{Krasnikov:1987jj} with bulk moduli and double
gaugino condensations.
The K\"ahler potential and the superpotential are obtained 
\begin{eqnarray}
K = -2 \log({\cal V}_E) , \qquad 
W = W_0 + \sum_i^{h_{+}^{1,1}} A_i e^{-a_iT_i}-B_i e^{-b_iT_i} ,
\end{eqnarray}
where $W_0 < 1 $.
Here, we add the uplifting potential,
\begin{eqnarray}
V_{\rm lift} &=& \frac{\epsilon}{{\cal V}_E^{4/3}} 
, \qquad 
\epsilon \simeq 3   \frac{\langle |W|^2 \rangle }{\langle {\cal V}_E^{2/3}\rangle}.
\end{eqnarray}
Then one finds via SUSY condition $D_iW \sim \partial_i W \sim 0$
\begin{eqnarray}
T_i &\simeq & \frac{1}{a_i-b_i} \log\left( \frac{a_iA_i}{b_iB_i}
\right) , 
\nonumber \\
\frac{F^{T_i}}{T_i+\overline{T_i}} &\simeq &
\frac{m_{3/2}}{a_ib_iT_i^2} \simeq \frac{m_{3/2}^2}{m_{T_i}}, \qquad 
m_{T_i} \simeq a_ib_i(T_i+\overline{T_i})^2 m_{3/2}^2 .
\end{eqnarray}
If one tunes $W_0$ to obtain $\langle W \rangle \sim 0$, 
moduli masses become much heavier than the gravitino mass \cite{Kallosh:2004yh}.

\end{document}